\pdfoutput=1
\documentclass[useAMS,usenatbib,usegraphicx]{mn2e}
\usepackage[svgnames]{xcolor}
\usepackage{apjfonts}
\usepackage{amssymb}
\usepackage{amsmath}
\usepackage{ctable}
\usepackage{fixltx2e} 

\newcommand{\ba}{\begin{eqnarray}}
\newcommand{\ea}{\end{eqnarray}}
\newcommand{\be}{\begin{equation}}
\newcommand{\ee}{\end{equation}}

\newcommand{\Msun}{M_{\odot}}
\newcommand{\Zsun}{Z_{\odot}}
\newcommand{\Ms}{M_{\ast}}
\newcommand{\Zs}{Z_{\ast}}

\newcommand{\Zg}{Z_{\rm gas}}
\newcommand{\Mvir}{M_{\rm vir}}
\newcommand{\Rvir}{R_{\rm vir}}
\newcommand{\pc}{{\rm pc}}
\newcommand{\cm}{{\rm cm}}
\newcommand{\nc}{n_{\rm th}}

\title[Galaxy Mass-Metallicity Relation]
{The Origin and Evolution of the Galaxy Mass--Metallicity Relation}

\author[X. Ma et al.]{
  \parbox[t]{1.0\textwidth}{ 
   Xiangcheng Ma,$^1$\thanks{E-mail: xchma@caltech.edu}
   Philip F. Hopkins,$^1$
   Claude-Andr{\'e} Faucher-Gigu{\`e}re,$^2$
   Nick Zolman,$^1$
   Alexander L. Muratov,$^3$
   Du{\v s}an Kere{\v s}$^3$ and
   Eliot Quataert$^4$
  } \vspace*{6pt} \\
  $^1$ TAPIR, MC 350-17, California Institute of Technology, Pasadena, CA 91125, USA \\ 
  $^2$ Department of Physics and Astronomy and CIERA, Northwestern University, 2145 Sheridan Road, Evanston, IL 60208, USA \\
  $^3$ Department of Physics, Center for Astrophysics and Space Sciences, University of California at San Diego, 9500 Gilman Drive, La Jolla, CA 92093 \\
  $^4$ Department of Astronomy and Theoretical Astrophysics Center, University of California Berkeley, Berkeley, CA 94720 
}

\pagerange{\pageref{firstpage}--\pageref{lastpage}}
\date{Draft version \today}

\begin{document}
\maketitle
\label{firstpage}

\begin{abstract}
We use high-resolution cosmological zoom-in simulations from the Feedback in Realistic Environment (FIRE) project to study the galaxy mass--metallicity relations (MZR) from $z=0$--6. These simulations include explicit models of the multi-phase ISM, star formation, and stellar feedback. The simulations cover halo masses $M_{\rm halo}=10^9$--$10^{13}~\Msun$ and stellar masses $\Ms=10^4$--$10^{11}~\Msun$ at $z=0$ and have been shown to produce many observed galaxy properties from $z=0$--6. For the first time, our simulations agree reasonably well with the observed mass--metallicity relations at $z=0$--3 for a broad range of galaxy masses. We predict the evolution of the MZR from $z=0$--6, as $\log(\Zg/\Zsun) = {\rm 12 + \log(O/H) - 9.0} = 0.35~[\log(\Ms/\Msun)-10] + 0.93 \exp(-0.43z) - 1.05$ and $\log(\Zs/\Zsun) = {\rm [Fe/H]} + 0.2 = 0.40~[\log(\Ms/\Msun)-10] + 0.67 \exp(-0.50z) - 1.04$, for gas-phase and stellar metallicity, respectively. Our simulations suggest that the evolution of MZR is associated with the evolution of stellar/gas mass fractions at different redshifts, indicating the existence of a universal metallicity relation between stellar mass, gas mass, and metallicities. In our simulations, galaxies above $\Ms=10^6~\Msun$ are able to retain a large fraction of their metals inside the halo, because metal-rich winds fail to escape completely and are recycled into the galaxy. This resolves a long-standing discrepancy between ``sub-grid'' wind models (and semi-analytic models) and observations, where common sub-grid models {\it cannot} simultaneously reproduce the MZR and the stellar mass functions.
\end{abstract}

\begin{keywords}
galaxies: formation -- galaxies: evolution -- cosmology: theory 
\end{keywords}

\section{Introduction}
\label{sec:intro}
The galaxy mass--metallicity relation (MZR) is one of the most fundamental properties observed in galaxies. In the local universe, there is a tight correlation between galaxy stellar mass and gas-phase oxygen abundance for star-forming galaxies \citep[e.g.][]{tremonti.04.sdss}, with an intrinsic scatter of only 0.1 dex in log(O/H). This relation has been extended to local dwarf galaxies and found to be a uniform, tight correlation over five orders of magnitude in stellar mass, from $\Ms=10^6$--$10^{11}~\Msun$ \citep{lee.06.mzr}. Many different groups have confirmed the MZR to exist not only in the local universe but also at high redshifts up to $z\sim2.3$ \citep[e.g.][]{savaglio.05.mzr,erb.06.mzr,zahid.11.deep2,zahid.12.mzr,andrews.13.mzr,henry.13a.mzr,henry.13b.mzr,yabe.14.mzr,steidel.14.mosfire,sanders.14.mzr}. \citet{zahid.13.mzr} compiled a number of the observed MZR from $z=0$--2.3 and found that the MZR evolves with redshift, with higher metallicity at low redshift for a given stellar mass. The MZR is also found at $z\gtrsim3$ \citep[e.g.][]{maiolino.08.mzr,mannucci.09.mzr}, despite the fact that the results are obtained from very small samples.

Gas-phase metallicities represent the ``current'' state of chemical enrichment in the galaxies, while stellar metallicities reflect the ``time-averaged'' galactic metallicity across the whole star formation history. Similarly, an MZR is also found in stellar metallicities. \cite{gallazzi.05.sdss} derived the stellar metallicities for $\sim$44,000 galaxies from SDSS and found a tight correlation between stellar mass and stellar metallicity for galaxies of stellar masses $10^9$--$10^{12}~\Msun$. \citet{kirby.13.mzr} measured the metallicities of individual stars in a sample of dwarf galaxies within the Local Group and found the SDSS stellar MZR can be continually extended down to $10^3~\Msun$. Despite the fact that stellar metallicity is challenging to measure at high redshifts, the stellar MZR provides very important and complimentary insights on the chemical evolution of galaxies, especially for massive quiescent galaxies and satellite galaxies in the local group where the gas-phase metallicities are hard to measure due to their low gas content.

Simple analytic models of galactic chemical evolution, such as the ``closed-box'', ``leaky-box'', and ``accreting-box'' models \citep[e.g.][]{schmidt.63.chem,talbot.71.chem,searle.72.chem,edmunds.90}, are often quoted to illustrate the qualitative behavior of the MZR. More complicated models have also been developed to work in cosmological contexts and to connect gas inflows, outflows, and star formation to galactic chemical evolution \citep[e.g.][]{dalcanton.07.model,finlator.dave.08,dave.12.model,lilly.13.model,lu.15.ana}. These models indicate that the existence of MZR is the consequence of an interplay between star formation efficiency, metal loss from gas outflows, and gas recycling and accretion. For example, the stellar mass--halo mass relation \citep[e.g.][]{moster.13.msmh,behroozi.13.msmh} indicates that the star formation efficiency (fraction of baryons turned into stars) is lower in low-mass galaxies than in more massive galaxies, suggesting that low-mass galaxies should be less metal-enriched. Meanwhile, galactic winds are ubiquitous \citep[see e.g.][for a recent review]{veilleux.05.winds}, carrying metals away from galaxies. Low mass galaxies have shallow potential wells so they tend to lose a significant fraction of their gas and metals, while massive galaxies have potential wells deep enough to prevent material from escaping the galaxy \citep[e.g.][]{dekel.silk.86}. On the other hand, gas inflows bring the metal-poor gas in the galactic halo and/or in the intergalactic medium (IGM) inwards, diluting the metal content in the interstellar medium (ISM) and supplying new material for star formation \citep[e.g.][]{keres.05.accretion,cafg.11.flow}. During this process, a considerable fraction of the gas and metals that have been formerly ejected via outflows eventually come back to the galaxy \citep[e.g.][]{bertone.07.recycle,oppenheimer.10.recycle}. Galaxy mergers and AGN activity could also be important, in the sense that they can trigger violent starburst, drive intensive gas outflows, and ultimately quench the star formation in the galaxy \citep[e.g.][]{springel.05.merger,hopkins.13.merger}.

Analytical models usually rely on simplified assumptions such as perfect mixing and adopt simple analytic prescriptions describing star formation, gas accretion, and outflows. In reality, these physical processes are tightly connected to each other and therefore must be treated self-consistently to understand the complete picture of galactic chemical evolution. Semi-analytic models (SAMs) of galaxy formation follow cosmological halo growth and halo mergers and include physically and/or empirically motivated prescriptions of heating and cooling, star formation, metal enrichment, gas accretion and outflows, recycling, and AGN feedback \citep[e.g.][]{croton.06.sam,somerville.08.sam,benson.12.sam,guo.13.sam,yates.13.sam,lu.11.sam,lu.13.sam,henriques.13.sam,henriques.15.sam,lu.15.3model}. They are much less computationally expensive to run than hydrodynamic simulations and are able to reproduce a number of galaxy properties for a broad range of stellar mass. However, one major challenge for SAMs is {\it simultaneously} reproducing observed stellar masses, star formation rates (SFRs), and metallicities. The metallicities of low-mass galaxies are particularly sensitive to the galactic wind model because strong outflows are required to suppress star formation in low-mass systems \citep[see e.g.][for a detailed comparison and discussion]{lu.13.sam}. Moreover, even though different SAMs have been succcessfully tuned to match the $z=0$ stellar mass function (SMF), many of them fail to match the observed the SMFs at high redshifts \citep[e.g.][]{somerville.15.araa}. At the same time, these models typically fail to match high-redshift MZR measurements and also diverge from one another in their MZR predictions.  
Nonetheless, it is encouraging that recently improved SAMs are able to reconcile stellar masses, colours, and SFRs of galaxies from $z=0$--3 \citep[e.g.][]{henriques.13.sam,henriques.15.sam}.

Large-volume cosmological hydrodynamic simulations produce large samples of galaxies and are powerful tools for statistical studies of galaxy properties \citep[e.g.][]{bertone.07.recycle,dave.11.mzr,torrey.14.illustris,schaye.15.eagle}. These simulations however usually have relatively poor mass and spatial resolution. They cannot explicitly resolve the multi-phase structure of the interstellar medium (ISM), when and where star formation takes place, how galactic winds are launched by stellar feedback, and how the winds interact with the circum-galactic medium (CGM). Approximate, empirical ``sub-grid'' models of the ISM structure, star formation, and stellar feedback are required and used. For example, \citet{dave.11.galaxy} implemented a momentum-driven wind model, with wind mass loading factors and velocities prescribed as a function of bulk galaxy properties. In their implementation, hydrodynamic interactions are temporarily suppressed as gas from the ISM is ``kicked'' into the galactic wind. Simulations using such simple prescriptions reveal similar problems to the SAMs. \citet{torrey.14.illustris} found a steeper slope than observed at the low-mass end of the MZR. These authors attributed it to the low metal retention efficiency in the presence of strong outflows, which were required in their model in order to prevent low-mass galaxies from forming too many stars. They further emphasized the tension between suppressing star formation and retaining enough metals in low-mass galaxies. Furthermore, the star formation histories in these simulations are very different and not all consistent with observations at high redshifts. Many cosmological simulations tend to form {\it too many} stars at early times (e.g. \citealt{dave.11.galaxy,sparre.14.illustris,fiacconi.15.argo}; for a review, see \citealt{somerville.15.araa}). Such problems are also common in SAMs. They are likely the result of imperfect star formation and stellar feedback models implemented in those simulations \citep[cf.][]{aquila.12}. Consequently, these simulations predict very different evolution of the MZR.  

Therefore, when using cosmological hydrodynamic simulations to understand the MZR and its evolution, one is required to capture the ``correct'' behavior of star formation, stellar feedback, gas outflows, and the mixing and interaction of galactic winds with the CGM on all relevant scales. Encouragingly, \citet{obreja.14.magicc} presented a suite of cosmological zoom-in simulations from the MaGICC project using an improved star formation and SNe feedback model. Their model includes an empirical prescription to approximate the effects of stellar feedback mechanisms operating before the first SNe explode. These authors showed that their simulations match the stellar mass--halo mass relation and the observed MZR from $z=0$--3, for the eight galaxies in their sample. In this work, the first of a series, we will study the chemical evolution of galaxies using the FIRE (Feedback in Realistic Environment) simulations \citep{hopkins.14.fire}. The FIRE project\footnote{FIRE project website: http://fire.northwestern.edu} is a series of cosmological zoom-in simulations that are able to follow galaxy merger history, interactions of galaxies with the IGM, and many other important processes. These simulations include a full set of realistic physical models and explicitly resolve the multi-phase structure of the ISM, star formation, and stellar feedback, with {\it no} need to tune parameters. The FIRE simulations successfully reproduce many observed galaxy properties, including the stellar mass-halo mass relation, star formation histories, the Kennicutt-Schmidt law, and the star-forming main sequence, from $z=0$--6, for a broad range of galaxy masses in $\Ms=10^4$--$10^{11}~\Msun$ \citep{hopkins.14.fire}. Also, the FIRE simulations predict reasonable covering fractions of neutral hydrogen in the halos of $z=2$--3 Lyman Break Galaxies \citep[LBGs;][]{cafg.14.fire} and self-consistently generate galactic winds with velocities and mass loading factors broadly consistent with observational requirements \citep{muratov.15.outflow}. These results further justify the reliability to study galactic chemical evolution using the FIRE simulations.

This paper focuses on the galaxy mass--metallicity relation. In companion papers, we will also study the stellar metallicity distribution functions and [$\alpha$/Fe] abundance ratio variation in dwarf galaxies, metallicity gradients and their origins, metal outflows and recycling. We start by describing the simulations in Section \ref{sec:simulations} and present the mass--metallicity relation at different redshifts in Section \ref{sec:mzr}. In Section \ref{sec:discussion}, we discuss the key processes that drive the shape and evolution of the MZR. We summarize and conclude in Section \ref{sec:conclusion}.

\begin{table*}
\caption{Simulation Initial Conditions.}
\centering
\begin{tabular}{lccccccc}
\hline\hline
Name & $M_{\rm halo}$ & $m_b$ & $\epsilon_b$ & $m_{\rm dm}$ & $\epsilon_{\rm dm}$ & Merger & Notes \\
 & ($\Msun$) & ($\Msun$) & ($\pc$) & ($\Msun$) & ($\pc$) & History & \\ 
\hline
{\bf m09} & 2.5e9 & 2.6e2 & 1.4 & 1.3e3 & 30 & normal & isolated dwarf \\
{\bf m10} & 0.8e10 & 2.6e2 & 3.0 & 1.3e3 & 30 & normal & isolated dwarf \\
{\bf m10lr} & 0.8e10 & 2.1e3 & 2.1 & 1.0e4 & 35 & normal & low resolution \\
{\bf m10v} & 0.8e10 & 2.1e3 & 7.0 & 1.0e4 & 70 & violent & -- \\
{\bf m11} &  1.4e11 & 7.0e3 & 7.0 & 3.5e4 & 70 & quiescent & -- \\ 
{\bf m11v} & 3.3e11 & 5.6e4 & 7.0 & 3.0e5 & 140 & violent & -- \\
{\bf m12v} &  6.3e11 & 3.9e4 & 10 & 2.0e5 & 140 & violent & several $z<2$ mergers \\ 
{\bf m12q} & 1.2e12 & 7.1e3 & 10 & 2.8e5 & 140 & late merger & -- \\ 
{\bf m12i} & 1.1e12 & 5.0e4 & 14 & 2.8e5 & 140 & normal & large ($\sim10\,R_{\rm vir}$) box \\ 
{\bf m13} &  6.0e12 & 3.6e5 & 21 & 2.2e6 & 210 & normal & small group mass \\
\hline
{\bf z2h350} & 7.9e11 & 5.9e4 & 9 & 2.9e5 & 143 & normal & -- \\
{\bf z2h400} & 7.9e11 & 5.9e4 & 9 & 2.9e5 & 143 & quiescent & -- \\
{\bf z2h450} & 8.7e11 & 5.9e4 & 9 & 2.9e5 & 143 & normal & -- \\
{\bf z2h506} & 1.2e12 & 5.9e4 & 9 & 2.9e5 & 143 & violent & -- \\
{\bf z2h550} & 1.9e11 & 5.9e4 & 9 & 2.9e5 & 143 & quiescent & -- \\
{\bf z2h600} & 6.7e11 & 5.9e4 & 9 & 2.9e5 & 143 & violent & -- \\
{\bf z2h650} & 4.0e11 & 5.9e4 & 9 & 2.9e5 & 143 & normal & -- \\
{\bf z2h830} & 5.4e11 & 5.9e4 & 9 & 2.9e5 & 143 & normal & -- \\
\hline
{\bf z5m09} & 7.6e8 & 16.8 & 0.14 & 81.9 & 5.6 & quiescent & ultra-high resolution \\
{\bf z5m10} & 1.3e10 & 131.6 & 0.4 & 655.6 & 7 & normal & ultra-high resolution \\
{\bf z5m10mr} & 1.4e10 & 1.1e3 & 1.9 & 5.2e3 & 14 & normal & -- \\
{\bf z5m11} & 5.6e10 & 2.1e3 & 4.2 & 1.0e4 & 14 & normal & -- \\
\hline\hline
\multicolumn{8}{p{0.65\linewidth}}{Parameters describing the initial conditions for our simulations (units are physical):} \\
\multicolumn{8}{p{0.65\linewidth}}{{\bf (1)} Name: Simulation designation.} \\
\multicolumn{8}{p{0.65\linewidth}}{{\bf (2)} $M_{\rm halo}$: Approximate mass of the main halo at $z=0$ ({\bf mxx} series), $z=2$ ({\bf z2hxxx} series), or $z=6$ ({\bf z5mxx} series). } \\
\multicolumn{8}{p{0.65\linewidth}}{{\bf (3)} $m_b$: Initial baryonic (gas and star) particle mass in the high-resolution region.} \\ 
\multicolumn{8}{p{0.65\linewidth}}{{\bf (4)} $\epsilon_b$: Minimum baryonic force softening (minimum SPH smoothing lengths are comparable or smaller. Force softening is adaptive (mass resolution is fixed).}\\
\multicolumn{8}{p{0.65\linewidth}}{{\bf (5)} $m_{\rm dm}$: Dark matter particle mass in the high-resolution region.} \\ 
\multicolumn{8}{p{0.65\linewidth}}{{\bf (6)} $\epsilon_{\rm dm}$: Minimum dark matter force softening (fixed in physical units at all redshifts).} 
\end{tabular}
\label{tbl:sim}
\end{table*}%

\section{The Simulations}
\label{sec:simulations}
\subsection{Simulation Details}
This work is part of the FIRE project. A full description of the numerical methods and physics included in our simulations is presented in \citep[][and references therein]{hopkins.14.fire}. We summarized their main features here. All the simulations use the newly developed GIZMO code \citep{hopkins.14.gizmo} in ``P-SPH'' mode. P-SPH adopts a Lagrangian pressure-entropy formulation of the smoothed particle hydrodynamics (SPH) equations \citep{hopkins.13.sph}, which eliminates the major differences between SPH, moving-mesh, and grid codes, and resolves many well-known issues in traditional density-based SPH formulations. The gravity solver is a heavily modified version of the GADGET-3 code \citep{springel.05.gadget}; and P-SPH also includes substantial improvements in the artificial viscosity, entropy diffusion, adaptive time-stepping, smoothing kernel, and gravitational softening algorithm.

We use the multi-scale ``zoom-in'' initial conditions generated with the MUSIC code \citep{hahn.11.music}, using second-order Lagrangian perturbation theory. The first set of simulations have been run down to $z=0$ and cover halo masses $10^9$--$10^{13}\Msun$ and stellar masses $10^4$--$10^{11}~\Msun$ at $z=0$ ({\bf mxx} series). Most of them have been presented in \citet{hopkins.14.fire}. The simulations {\bf m09} and {\bf m10} are isolated dwarfs, constructed using the method from \citet{onorbe.14.dwarf}. Simulations {\bf m10v}, {\bf m11}, {\bf m11v}, {\bf m12q}, {\bf m12i}, and {\bf m13} are chosen to match the initial conditions from the AGORA project \citep{kim.14.agora}, which will enable future comparisons with a wide range of simulation codes and physics implementations. Simulation {\bf m12v} is based on the initial conditions studied in \citet{keres.09.b1ic} and \citet{cafg.11.b1ic}. The simulations with a label `{\bf v}' have relatively violent merger histories at $z<2$, while the rest have more typical merger histories. The resolution of these simulations is chosen to scale with the mass of the system to ensure we are able to resolve the giant molecular clouds (GMCs). We also include a separate set of simulations run to $z=2$ ({\bf z2hxxx} series), which are presented in \citep{cafg.14.fire}. Their main halos are chosen to host Lyman break galaxies (LBG) and cover halo masses $1.9\times10^{11}$--$1.2\times10^{12}\Msun$ at $z=2$. Finally, we include another series of simulations only run to $z\sim6$, but with extremely high resolutions ({\bf z5mxx} series). These simulations are presented in \citet{ma.15.escape}. Their main halos cover halo masses from $7.7\times10^8$--$5.6\times10^{10}\Msun$ at $z=6$ and these galaxies are believed to contribute most to the cosmic reionization \citep[e.g.][]{kuhlen.fg.12,robertson.13.udf12}. The initial conditions of all the simulations are summarized in Table~\ref{tbl:sim}.

In our simulations, gas follows an ionized+atomic+molecular cooling curve from 10--$10^{10}$ K, including metallicity-dependent fine-structure and molecular cooling at low temperatures and high-temperature metal-line cooling followed species-by-species for 11 separately tracked species \citep[H, He, C, N, O, Ne, Mg, Si, S, Ca, and Fe; see][]{wiersma.09.cooling}. At each timestep, the ionization states and cooling rates are determined from a compilation of CLOUDY runs, including a uniform but redshift-dependent photo-ionizing background tabulated in \citet{fg.09.uvb}, and photo-ionizing and photo-electric heating from local sources. Gas self-shielding is accounted for with a local Jeans-length approximation, which is consistent with the radiative transfer calculations in \citet{fg.10.lya}.

Star formation is allowed only in dense, molecular, and self-gravitating regions with hydrogen number density above some threshold $\nc=10$--100 $\cm^{-3}$ \citep{hopkins.13.sf}. Stars form at 100\% efficiency per free-fall time when the gas meets these criteria. The self-gravity criterion is physically required to obtain the correct spatial star formation distribution in galaxies \citep{hopkins.13.sf,padoan.11.sf}, but the galaxy-averaged star formation efficiency is regulated by feedback at much lower values \citep[$\sim1$\% per dynamical time, e.g.][]{cafg.13.feedback}. We stress that changing these parameters in a reasonable range only yields small and random variations to the global star formation history, as long as feedback is active \citep[see][]{hopkins.11.fb,hopkins.12.fb}. 

Once a star forms, it inherits the metallicity of each tracked species from its parent gas particle. Every star particle is treated as a single stellar population with known mass, age, and metallicity. Then all the feedback quantities, including ionizing photon budgets, luminosities, stellar spectra, supernovae (SNe) rates, mechanical luminosities of stellar winds, metal yields, etc., are directly tabulated from the stellar population models in STARBURST99 \citep{leitherer.99.sb}, assuming a \citet{kroupa.02.imf} initial mass function (IMF) from $0.1$--$100~\Msun$\footnote{In principle, the ``IMF-averaged'' approximation does not hold for the ultra-high resolution simulations in the {\bf z5mxx} series, where the mass of a star particle is only $10$--$100~\Msun$. Nevertheless, we confirmed that these simulations predict similar global galaxy properties to those of much poorer resolutions \citep[see][]{ma.15.escape}.}. We account for several different stellar feedback mechanisms, including: {\bf (1)} local and long-range momentum flux from radiative pressure; {\bf (2)} energy, momentum, mass and metal injection from SNe and stellar winds; and {\bf (3)} photo-ionization and photo-electric heating. We follow \citet{wiersma.09.metal} and include the metal yields from Type-II SNe, Type-I SNe, and stellar winds. We note that the Type-II SNe yield table from \citet{woosley.95.metal} adopted in our simulations produce Mg roughly $\sim0.4$ dex below the typical values in modern models \citep[e.g.][]{nomoto.06}. This will have little effect on the galaxy properties in our simulations, as Mg is not an important coolant. Nevertheless, we will add 0.4 dex to the Mg abundance to correct this in the analysis below.

All simulations adopt a standard flat $\Lambda$CDM cosmology with cosmological parameters consistent with $H_0=70.2 {\rm~km~s^{-1}~Mpc^{-1}}$, $\Omega_{\Lambda}=0.728$, $\Omega_{m}=1-\Omega_{\Lambda}=0.272$, $\Omega_b=0.0455$, $\sigma_8=0.807$ and $n=0.961$ \citep[e.g.][]{hinshaw.13.wmap,planck.13}.

\begin{figure}
\centering
\includegraphics[width=0.5\textwidth]{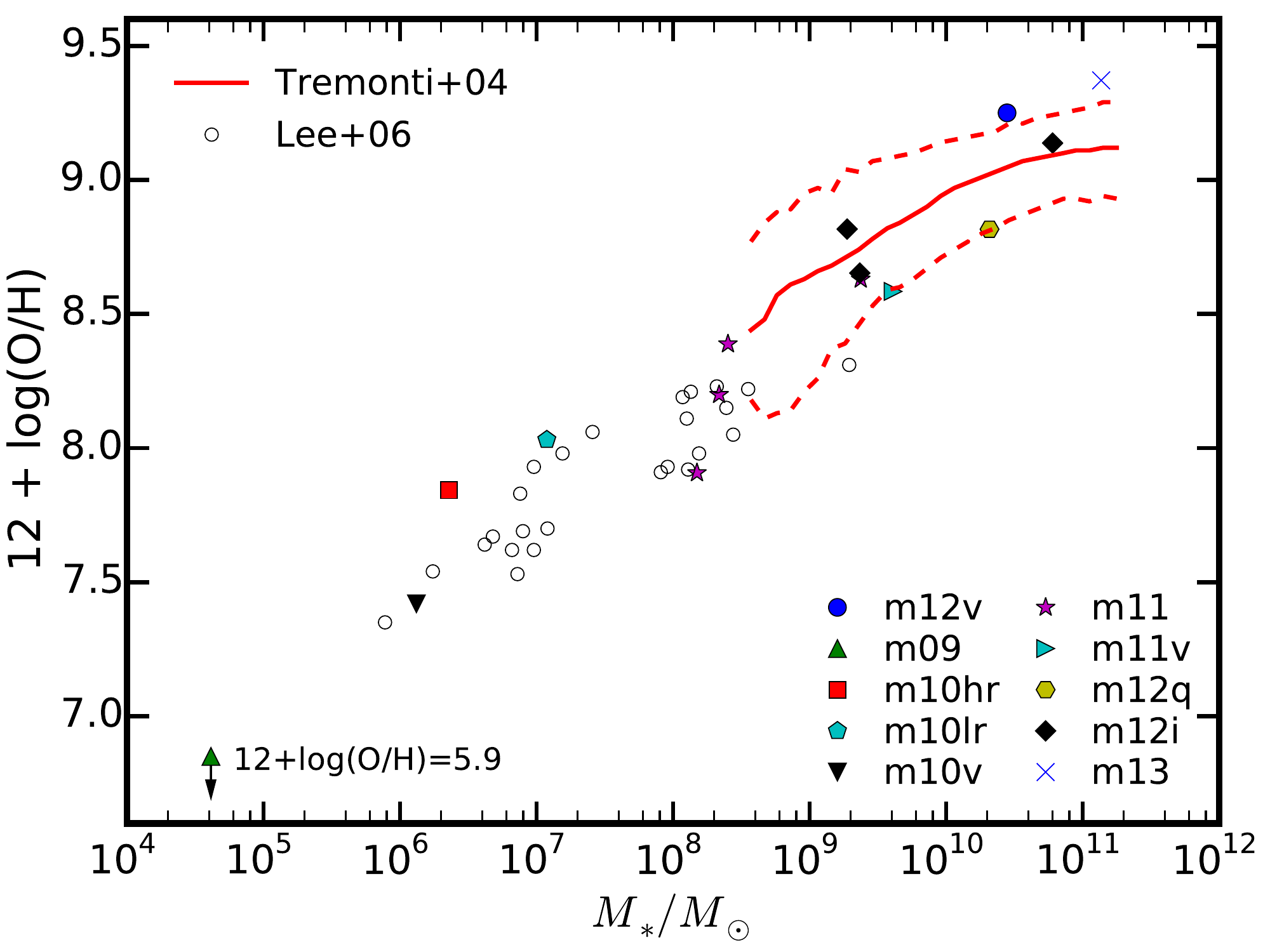}
\caption{Stellar mass--gas-phase oxygen abundance relation at $z=0$. The red solid and dashed curves represent the median and $2\sigma$ dispersion of the SDSS MZR at $z\sim0.1$ \citep{tremonti.04.sdss}. The open circles denote the data of the dwarf galaxy sample from \citet{lee.06.mzr}. Our simulations are in good agreement with observations from $\Ms=10^6$--$10^{11}~\Msun$.}
\label{fig:MZRz0Gas}
\end{figure}

\begin{figure}
\centering
\includegraphics[width=0.5\textwidth]{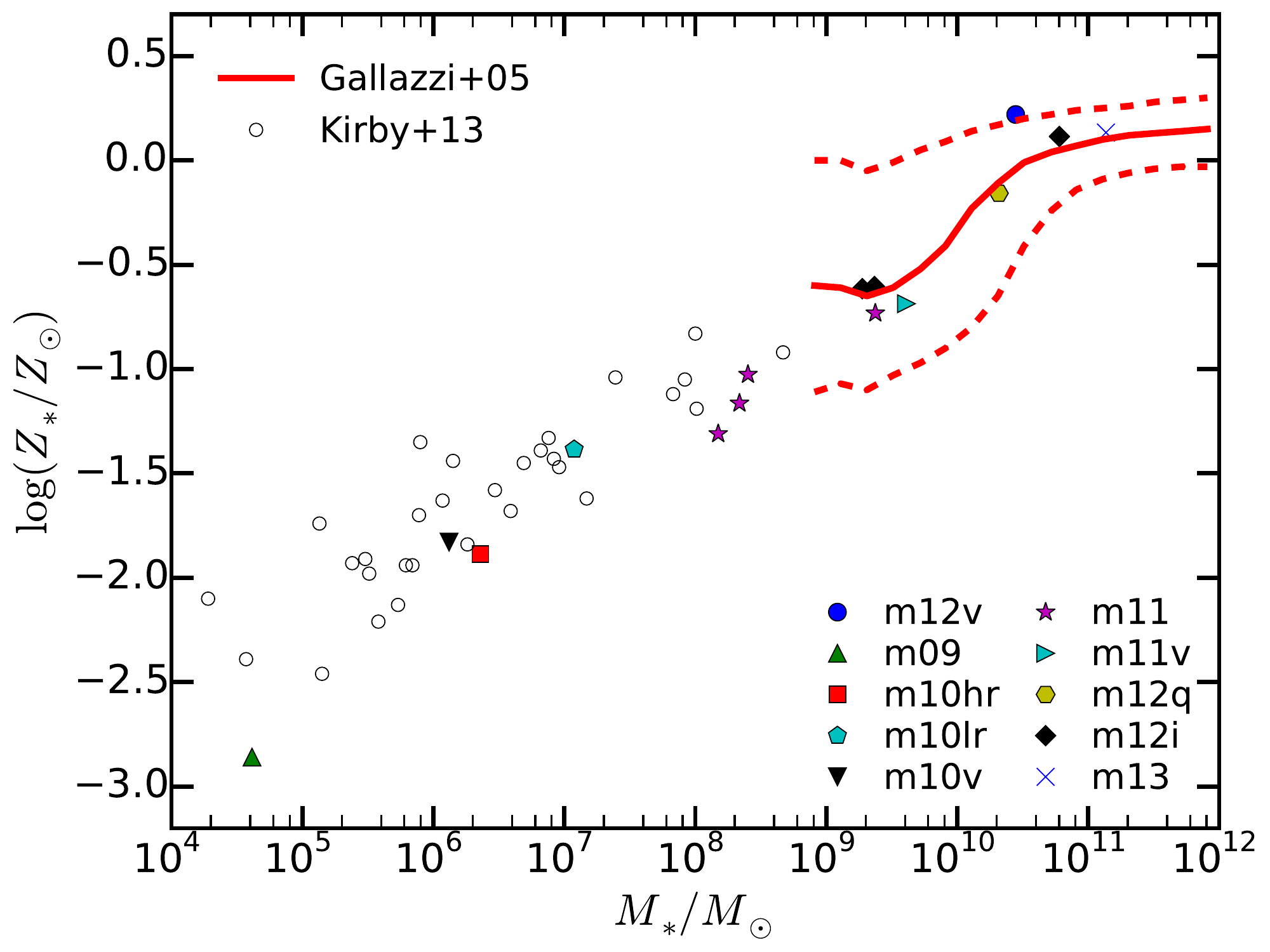}
\caption{Stellar mass--stellar metallicity relation at $z=0$. The red solid and dashed curves are the median and $1\sigma$ dispersion of the SDSS MZR in the local universe \citep{gallazzi.05.sdss}. The open circles represent the values of [Fe/H] of individual dwarfs from \citet{kirby.13.mzr}. Again, the agreement is good from $10^4$--$10^{11}~\Msun$.}
\label{fig:MZRz0Star}
\end{figure}

\begin{figure*}
\centering
\includegraphics[width=0.95\textwidth]{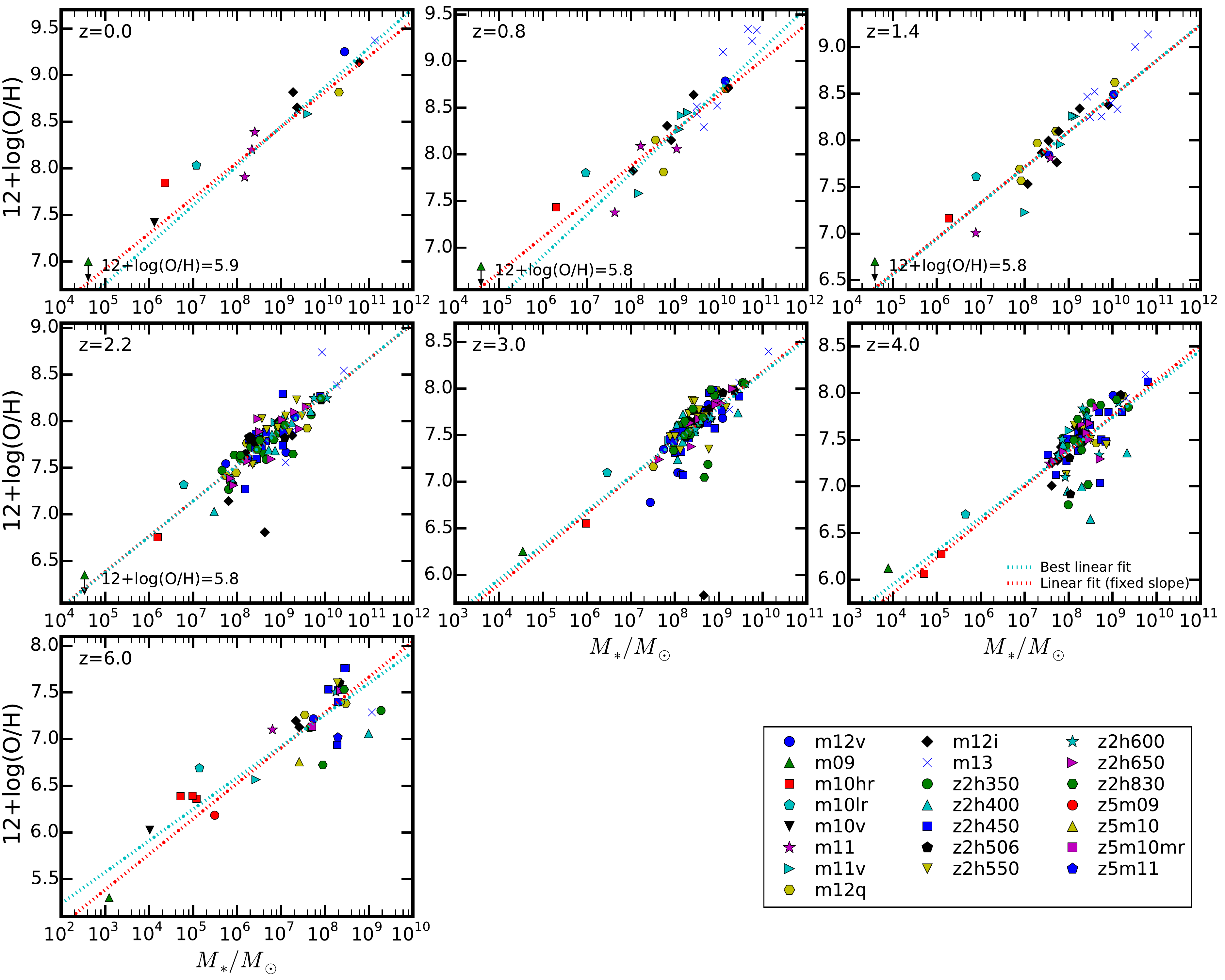}
\caption{Stellar mass--gas-phase metallicity relation at all redshifts. Cyan dotted lines show the best linear fit $\log(\Zg/\Zsun) = {\rm 12+\log(O/H)} - 9.0 = \gamma_g [\log(\Ms/\Msun)-10] + Z_{g,10}$. The red dotted lines show the best fit for a fixed slope $\gamma_g=0.35$. Note that a constant slope provides a very good fit, where the zero point evolves by $\sim1$ dex from $z=0$--6.}
\label{fig:MZRGasAllz}
\end{figure*}

\begin{figure*}
\centering
\includegraphics[width=0.95\textwidth]{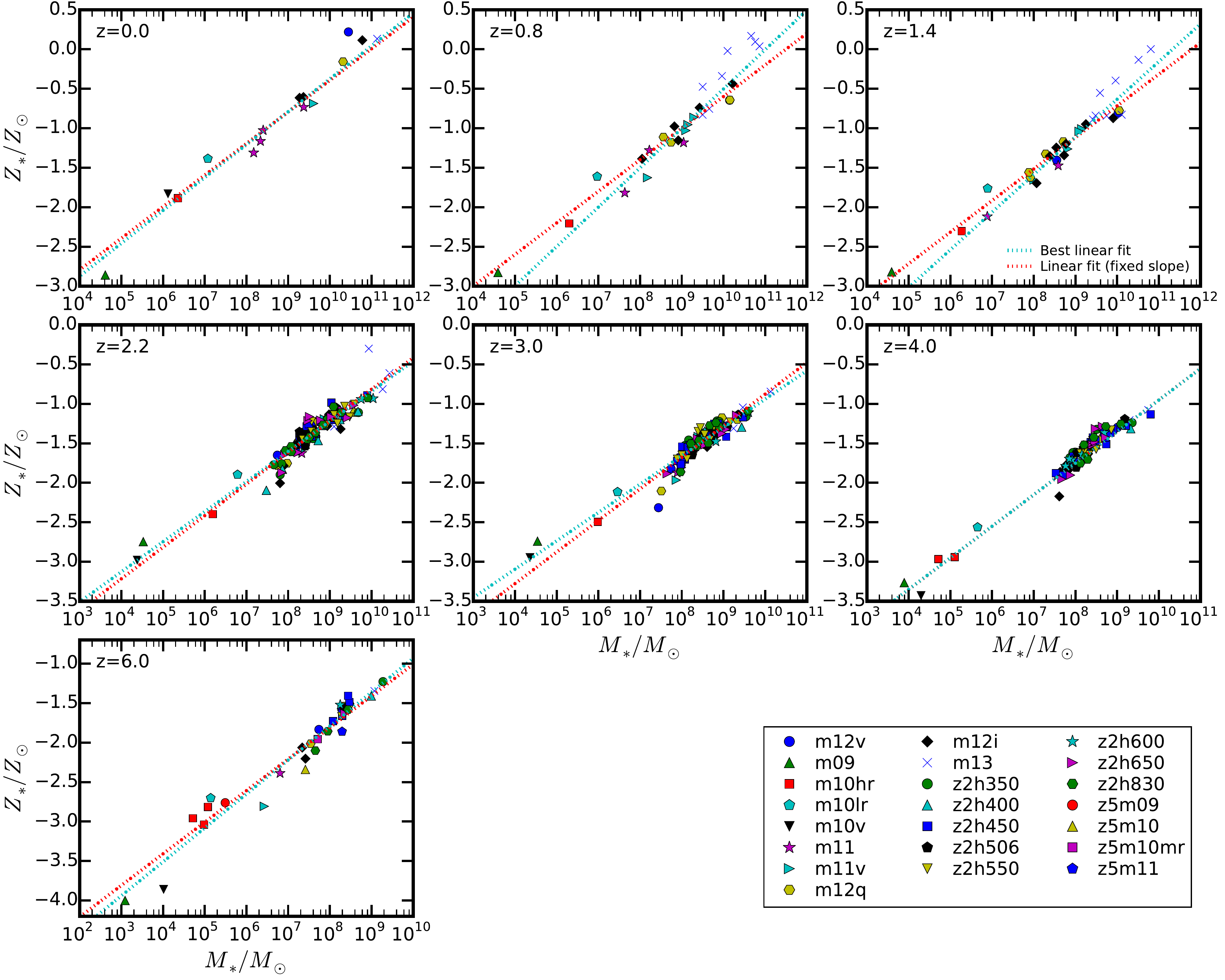}
\caption{Stellar mass--stellar metallicity relation at all redshifts. Cyan dotted lines show the best linear fit at each redshift $\log(\Zs/\Zsun) = {\rm [Fe/H] +0.2} = \gamma_{\ast} [\log(\Ms/\Msun)-10]  + Z_{\ast,10}$. The red dotted lines show the best fit for a fixed slope $\gamma_{\ast}=0.40$. Again, the slope is approximately constant, while the normalization evolves by $\sim1$ dex.}
\label{fig:MZRStarAllz}
\end{figure*}

\begin{figure}
\centering
\includegraphics[width=0.5\textwidth]{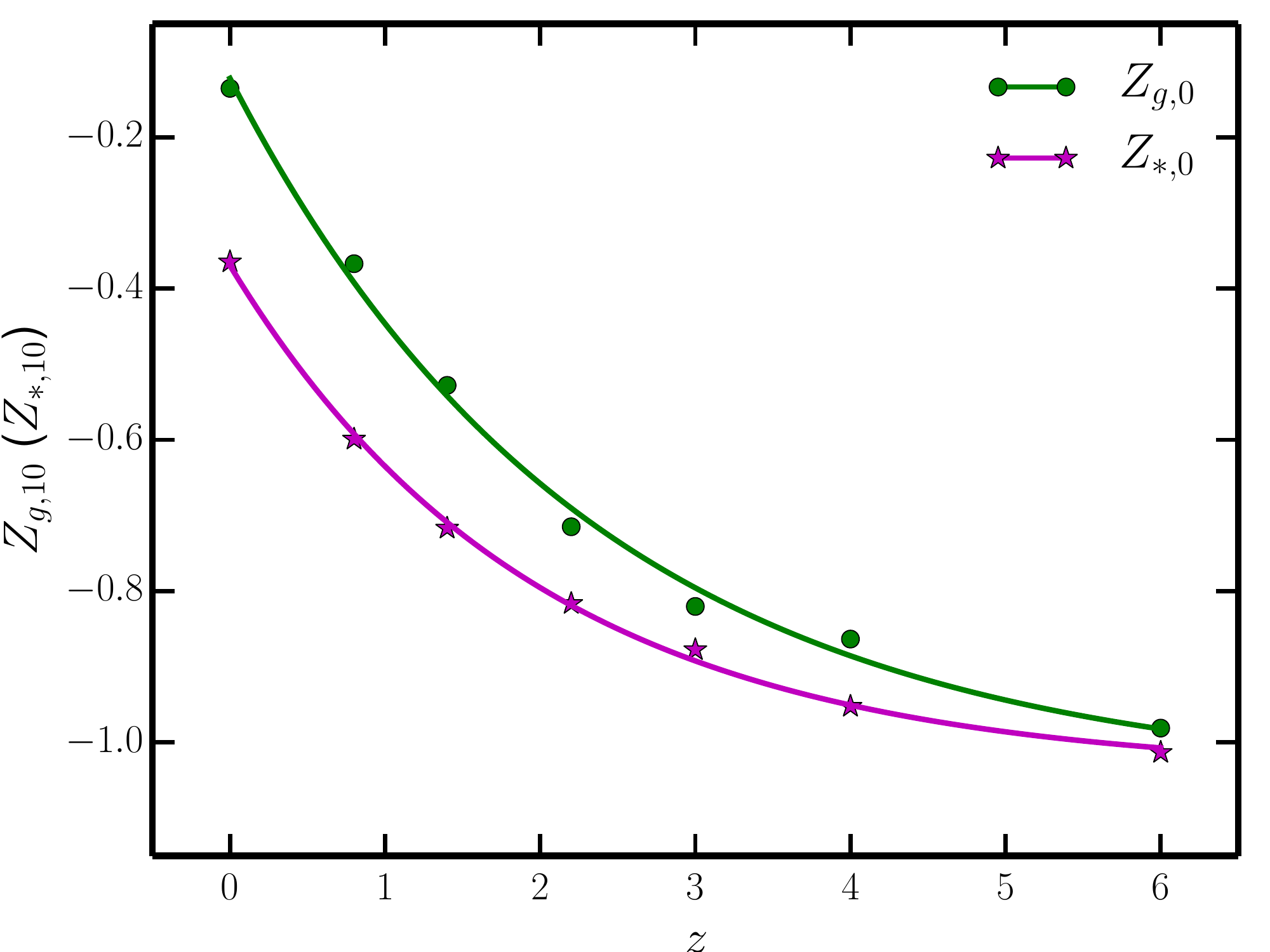}
\caption{The gas-phase and stellar metallicity at $\Ms=10^{10}~\Msun$, $Z_{g,10}$ and $Z_{\ast,10}$ as a function of redshift. The solid lines are the best fit of exponential functions $Z_{g,10} = 0.93 \exp(-0.43z) - 1.05$ and $Z_{\ast,10}= 0.67 \exp(-0.50z) - 1.04$.}
\label{fig:ZeroPoint}
\end{figure}

\begin{figure*}
\centering
\includegraphics[width=0.75\textwidth]{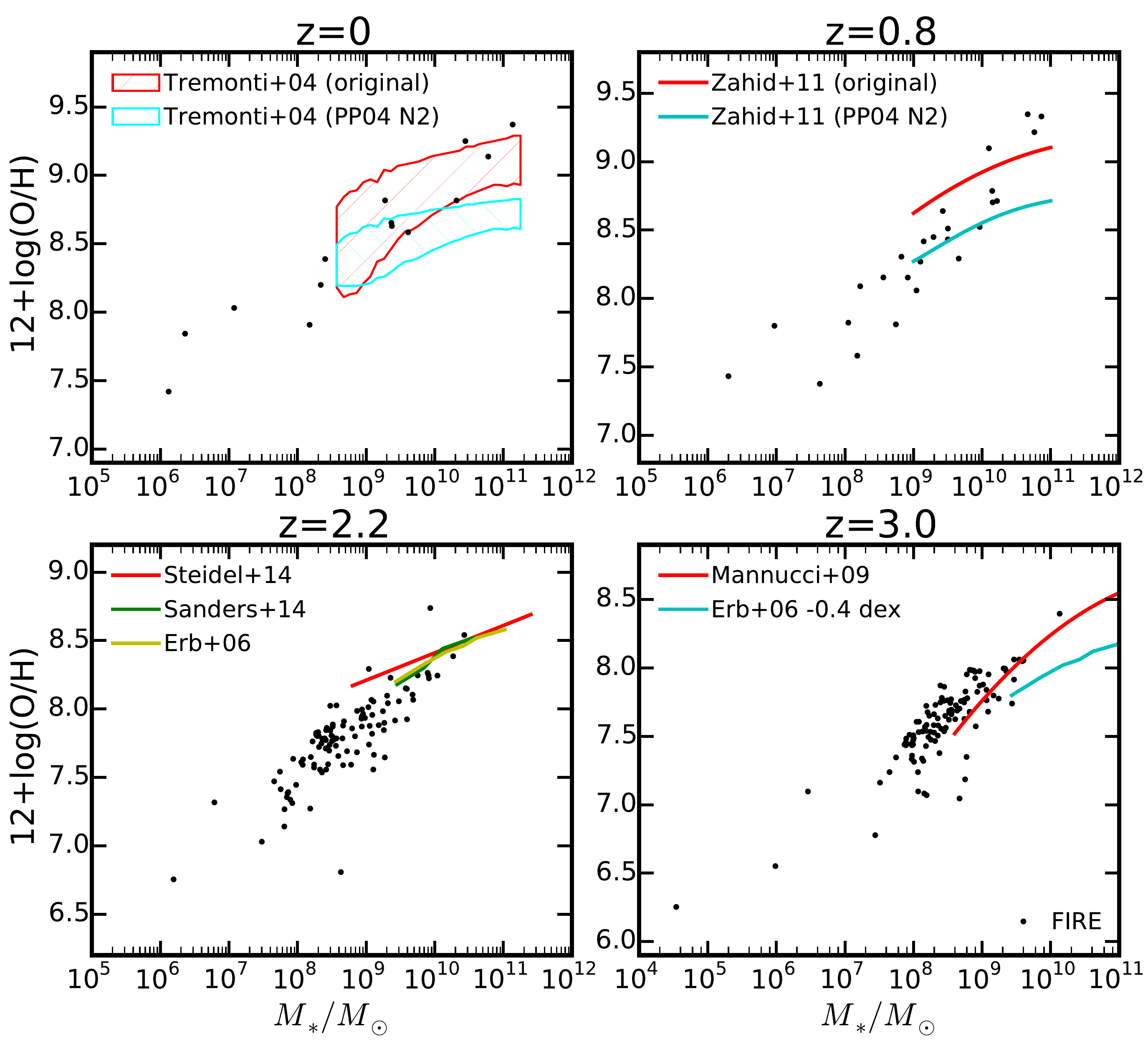}
\caption{Stellar mass--gas-phase oxygen abundance relations at $z=0$, 0.8, 2.2, and 3.0, as compared with a number of observations at these redshifts. In the upper panels, we show both the original relations (red lines) from \citet[][$z\sim0$]{tremonti.04.sdss} and \citet[][$z\sim0.8$]{zahid.11.deep2} and the relations converted to PP04 N2 calibration (cyan lines) following \citet{kewley.ellison.08}. In the lower left panel, we show the observed MZR at $z\sim2.3$ from \citet[][the red line]{steidel.14.mosfire}, \citet[][the green line]{sanders.14.mzr}, and \citet[][the yellow line]{erb.06.mzr}. In the lower right panel, we show the best fitting from \citet[][$z\sim3.1$]{mannucci.09.mzr}. We also shift the \citet{erb.06.mzr} data downward by 0.4 dex for a comparison as motivated by Figure 5 in \citet{mannucci.09.mzr}. Our simulations are broadly consistent with observations over a wide range of stellar mass from $z=0$--3, given the significant systematic uncertainties observational determinations of metallicities.}
\label{fig:MZRGasObsComp}
\end{figure*}

\begin{figure*}
\centering
\includegraphics[width=0.75\textwidth]{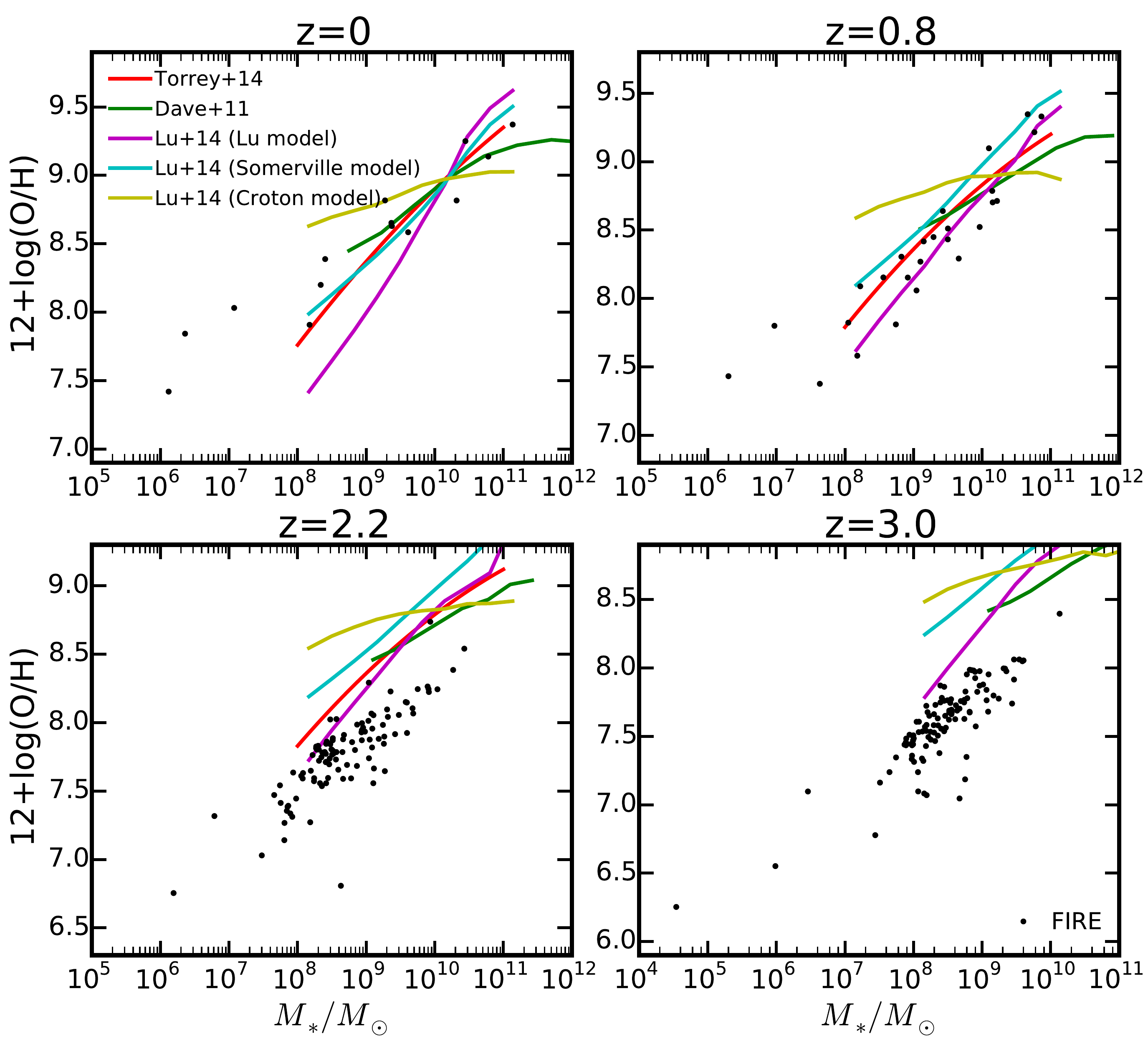}
\caption{Stellar mass--gas-phase oxygen abundance relation at $z=0$, 0.8, 2.2, and 3.0, as compared with other numerical simulations and semi-analytic models. We renormalize other works to ${\rm 12+log(O/H)=8.9}$ at $\Ms=10^{10}\Msun$ at $z=0$ with respect to our simulations. Red and green lines show the results from cosmological simulations presented in \citet{torrey.14.illustris} and \citet{dave.11.mzr}, respectively, which used popular ``sub-grid'' models for galactic winds. Magenta, cyan, and yellow lines show the predictions of three semi-analytic models from \citet[][the Lu model, the Somerville model, and the Croton model, respectively]{lu.13.sam}. All of these models reproduce the correct $z=0$ stellar mass function, but none of them correctly reproduces the slope or the redshift evolution of the MZR.}
\label{fig:MZRGasTheComp}
\end{figure*}

\subsection{Halo Identification, Stellar Mass and Metallicity}
\label{sec:halo}
We use the Amiga Halo Finder \citep[AHF;][]{gill.04.ahf,knollmann.09.ahf} to identify galactic halos and galaxies in our simulations. The AHF code uses the adaptive mesh refinement method and identifies halos and subhalos as groups of {\it bound} particles (dark matter, gas, and stars). In this work, we only consider those ``well-resolved'' halos that include more than $10^5$ bound particles, have at most $10\%$ of their mass contaminated by low-resolution particles, and contain at least 100 gas and 100 star particles, respectively. These criteria are somewhat arbitrary; but varying these numbers within a reasonable range will have little effect on our conclusions. If none of the halos meets these criteria in a snapshot (this happens in some snapshots at high redshifts ($z\sim6$), where the galaxy progenitors are too small to contain so many particles), we will take the most massive halo in the high-resolution region in our analysis. We do not include subhalos/satellite galaxies in this work. The centre of a halo is located at the centre of mass of the finest refinement level. We adopt the virial overdensity from \citet{bryan.norman.98}, which evolves with cosmic time.

We only consider the main galaxy in each halo. To remove the contamination of satellite galaxies, we exclude any gas/star particle that is bound to a subhalo in the analysis below. We measure the galaxy stellar mass ($\Ms$) by summing over the mass of all star particles that belong to the main galaxy. Then we define its stellar metallicity (as well as the abundance of each tracked species) as mass-averaged metallicity (abundance) of all star particles. To separate halo gas and the ISM, we apply a simple temperature criteria and select all gas particles below $10^4$ K as the ISM. In our simulations, this is equivalent to selecting gas above some density threshold of a few $0.1~\cm^{-3}$ (we explicitly check the gas distribution in the density--temperature plane), which is comparable to the mean gas density within a few stellar effective radii. It naturally picks warm ionized and cold neutral gas. We define the gas-phase metallicity as the mass-weighted metallicity of all gas particles that belong to the ISM (we compare and discuss three different definitions of gas-phase metallicity in Appendix \ref{sec:zdefine})\footnote{In many cosmological simulations with ``sub-grid'' models, gas-phase metallicity is usually defined as star-formation-rate-averaged metallicity. However, our simulations explicitly resolve multi-phase ISM structures and include realistic models of star formation and feedback. Individual gas particles are very sensitive to local feedback processes. For these reasons, we do not apply SF-averaged gas-phase metallicity to our simulations.}.

In this work, we use $\Zg$ and $\Zs$ to refer to the mass fraction of all heavy elements in gas and stars, respectively. In Section~\ref{sec:mzr}, we will primarily use oxygen abundance $\rm 12+log(O/H)$ to present gas-phase metallicities, which is defined in terms of number ratio of oxygen to hydrogen atoms, in order to directly compare with observations. For stellar metallicity, we will primarily use $\Zs$ in the rest of this work. In the literature, gas-phase metallicity and stellar metallicity are also sometimes referred as $\Zg$ and iron abundance [Fe/H] (in solar units), respectively. For these reasons, we provide the conversion between these quantities for our simulated galaxies. We will show the calibration in Appendix \ref{sec:zcalib} but directly give the results here. For a solar metallicity of 0.02 and a solar iron abundance 0.00173 (both in mass fraction), we obtain ${\rm 12+log(O/H)}=\log(\Zg/\Zsun)+9.00$ and ${\rm [Fe/H]}=\log(\Zs/\Zsun)-0.20$. We emphasize that these relations may suffer from systematic uncertainties that originate from: {\bf (1)} Type-II and Type-I SNe rates, {\bf (2)} metal yields of tracked species from different channels, and {\bf (3)} the solar abundances we adopt in our simulations. However, the shape and evolution of the MZR should be robust to these uncertainties.

\section{The Mass--Metallicity Relation}
\label{sec:mzr}
In this section, we present both the gas-phase and stellar MZR from $z=$0--6 and compare our results with observations and other simulations. We will further explore the most important factors that shape the MZR and drive its evolution in the Section~\ref{sec:discussion}.

\subsection{The MZR at $z=0$}
We begin by showing the gas-phase MZR at $z=0$. In Figure~\ref{fig:MZRz0Gas}, we present the stellar mass--gas-phase oxygen abundance relation for our {\bf mxx} series simulations at $z=0$. For comparison, we also present the median and $2\sigma$ dispersion of the SDSS MZR from \citet[][red solid and dashed lines]{tremonti.04.sdss} and the data of individual local dwarf galaxies compiled in \citet[][open circles]{lee.06.mzr} in Figure~\ref{fig:MZRz0Gas}. We remind the reader that these observed gas-phase oxygen abundances are derived from the relative strength of strong nebulae emission lines produced by photo-ionization from young massive stars, so that the observed gas-phase MZR only holds for star-forming galaxies. Also, we emphasize that the overall shape of gas-phase MZR strongly depends on which empirical calibration it uses and the normalization of this relation differs by at most 0.7 dex between different calibrations \citep[][see also Figure \ref{fig:MZRGasObsComp}]{kewley.ellison.08}. For these reasons, we do not apply any correction to these observed data but keep them in their original forms.

In general, our simulations agree reasonably well with observations across stellar mass from $\Ms=10^6$--$10^{11}~\Msun$. However, our simulations do not show evidence for flattening at the high-mass end of the gas-phase MZR. The gas-phase metallicity increases with stellar mass up to $\Ms\sim10^{11}~\Msun$ in our sample. The simulations predict slightly higher metallicities than the observed relation from \citet{tremonti.04.sdss} at $\Ms=10^{11}~\Msun$. The most significant discrepancy is due to our m13 run, which is a somewhat lower resolution simulation of a massive galaxy and which did not include the possible effects of AGN feedback. For example, as it has been shown in \citet{hopkins.14.fire}, the main galaxy in {\bf m13} have the cooling flow problem and never quenches at low redshift. The SFR of this galaxy is $3~\Msun$ yr$^{-1}$, which is fairly low in its star formation history, but significantly higher than observationally inferred values below $z\sim1$. Consequently, this galaxy might be over-enriched at low redshift. If so, this suggests that additional physics, such as AGN feedback, is probably required to fully understand the chemical evolution in massive galaxies, at least in the sense of quenching star formation. Alternatively, it has also been proposed that the observed MZR could continue to rise at the high-metallicity end when using new metallicity diagnostics that account for non-equilibrium electron energy distributions \citep[see e.g.][]{dopita.13.kappa,nicholls.13.kappa}. Furthermore, we note that the ``flatness'' of MZR at the high-mass end behaves very differently when applying different empirical calibrations \citep[e.g.][]{kewley.ellison.08}. Therefore, we do not further quantitatively discuss the discrepancy between our simulations and observations at the massive-end of MZR, but rather focus on galaxies below $\Ms=10^{11}~\Msun$ where our simulations are most robust. A larger sample of simulations with improved resolution at the massive end is required to make a robust comparison.

Most of our simulated galaxies are still forming stars (at least very weakly) at $z=0$, except for {\bf m09}. The {\bf m09} is a low-mass isolated dwarf galaxy (comparable to the ultra faint dwarfs around the Milky Way), in which star formation has been shut down since $z=3$ by cosmic reionization \citep{onorbe.15.dwarf}. At $z=0$, this galaxy has lost almost all metals it produced (see Section \ref{sec:discussion}). Although its gas-phase metallicity is an order of magnitude lower than the extrapolation of the observed MZR down to $\Ms=10^4~\Msun$, it is not contradictory to observations in the sense that the gas-phase metallicity of such galaxies cannot be measured due to lack of strong nebular emission lines. 

In Figure~\ref{fig:MZRz0Star}, we show the stellar mass--stellar metallicity relationship at $z=0$ and compare our simulations with the SDSS sample from \citet[][red solid and dashed curves]{gallazzi.05.sdss} and the dwarf galaxies from \citet[][open circles]{kirby.13.mzr}. Note that the stellar metallicities from \citet{gallazzi.05.sdss} are measured from absorption features of galaxy-integrated spectra (mostly Mg and Fe lines), while the metallicities from \citet{kirby.13.mzr} are derived from Fe abundances of individual stars. The conversion between different methods and their systematic uncertainties are complex and beyond the scope of this paper. For our purposes, we avoid any correction to these observations but present them in their original values\footnote{In Figure \ref{fig:MZRz0Star}, we plot the values of [Fe/H] from \citet{kirby.13.mzr}, avoiding the complicated conversion between [Fe/H] and $\Zs/\Zsun$ for the observed sample.}.

Our simulations match these observations quite well over the whole mass coverage from $\Ms=10^4$--$10^{11}~\Msun$. The simulated sample shows a flatness in the stellar MZR around $\Ms=10^{11}~\Msun$ at $z=0$, consistent with the observed SDSS MZR from \citet{gallazzi.05.sdss}. This is the consequence of the fact that the growth of the more massive galaxies in our simulations is dominated by mergers and accretion of low-mass metal-poor satellites rather than {\it in situ} star formation at low redshifts. Therefore, the average stellar metallicities do not strongly evolve despite the fact that the stellar masses may grow considerably at low redshifts \citep[see also][]{choi.14.mzr}.

\subsection{Evolution of the MZR}
\label{sec:mzrevolve}
Figure \ref{fig:MZRGasAllz} and \ref{fig:MZRStarAllz} show the gas-phase and stellar MZR, respectively, from $z=0$--6. We note that for $z\gtrsim2$ and $z=6$, we include the {\bf z2hxxx} and {\bf z5mxx} simulations in our analysis. The stellar MZR is tighter than the gas-phase MZR, i.e., the gas-phase MZR has larger scatter than stellar MZR at fixed stellar mass. This is because in our simulations, especially at high redshifts, star formation is dominated by multiple bursts, which drives bursts of gas outflows \citep{muratov.15.outflow}. As a consequence, instantaneous gas-phase metallicities may have considerable time fluctuations associated with gas inflows, outflows, and mergers. This effect is larger at high redshifts when the galaxy progenitors are of much lower masses and galaxy mergers are more common, resulting in some outliers that deviate from the main MZR at high redshifts. Despite the short-time-scale fluctuations, both the gas-phase and stellar metallicities increase with time on cosmological time-scales. At all times, gas-phase metallicities are higher than stellar metallicities, since gas-phase metallicities represent the current state of metal enrichment in the galaxies, while stellar metallicities reflect the average galactic metallicities across the whole time. Both metallicities should converge at high redshifts.

To illustrate this quantitively, we fit the gas-phase and stellar MZR at different redshifts for our simulated galaxies with simple linear functions $\log(\Zg/\Zsun) = {\rm 12+\log(O/H)} - 9.0 = \gamma_g [\log(\Ms/\Msun)-10] + Z_{g,10}$ and $\log(\Zs/\Zsun) = {\rm [Fe/H] +0.2} = \gamma_{\ast} [\log(\Ms/\Msun)-10]  + Z_{\ast,10}$, where $\gamma_g$ and $\gamma_{\ast}$ are the slopes and $Z_{g,10}$ and $Z_{\ast,10}$ represent the typical gas-phase metallicity and stellar metallicity at $\Ms=10^{10}~\Msun$. Although simple linear function do not capture the flatness of stellar metallicity above $\Ms\sim10^{11}~\Msun$ at $z<1$, it is sufficient for our purposes here. We use least-squares fitting to obtain the best fit (the cyan dotted lines in Figure \ref{fig:MZRGasAllz} and \ref{fig:MZRStarAllz}). In principle, both the slopes and zero points should be functions of redshift. Nevertheless, the MZR at different redshifts have very similar slopes. For simplicity, we pick the mean slope of each relation and redo the linear fit using fixed slopes. We choose $\gamma_g=0.35$ and $\gamma_{\ast}=0.40$ (red dotted lines in Figure \ref{fig:MZRGasAllz} and \ref{fig:MZRStarAllz}) and confirm that both the best linear fit and the fixed-slope fit describe the simulations reasonably well. We then attribute the evolution of MZR to the evolution of $Z_{g,10}$ and $Z_{\ast,10}$ with redshift, which we show in Figure~\ref{fig:ZeroPoint}. We fit these parameters as a function of redshift by an exponential function $F(z) = A \exp(-Bz) + C$. The best fit gives $Z_{g,10} = 0.93 \exp(-0.43z) - 1.05$ and $Z_{\ast,10}= 0.67 \exp(-0.50z) - 1.04$, respectively (the green and magenta lines in Figure~\ref{fig:ZeroPoint}). These give the gas-phase and stellar MZR from $z=0$--6 as $\log(\Zg/\Zsun) = {\rm 12 + \log(O/H) - 9.0} = 0.35~[\log(\Ms/\Msun)-10] + 0.93 \exp(-0.43z) - 1.05$ and $\log(\Zs/\Zsun) = {\rm [Fe/H]} + 0.2 = 0.40~[\log(\Ms/\Msun)-10] + 0.67 \exp(-0.50z) - 1.04$, respectively.

In general, the fitting functions above represent the gas-phase and stellar MZR fairly well for our simulated galaxies, except for the flattening of the stellar MZR above $\Ms\sim10^{11}~\Msun$ at $z=0$. We emphasize that these results have systematic uncertainties from Type-II and Type-Ia SNe rates, the solar abundance, and the metal yield tables we implement in our simulations. When using these fitting functions, one should notice the uncertainties and make adjustments accordingly.

\subsection{Comparison with Observations and Other Models}
In Figure~\ref{fig:MZRGasObsComp}, we compare the gas-phase MZR between our simulations and a number of observations at multiple redshifts. We show the observed MZR at $z\sim0$ \citep{tremonti.04.sdss}, $z\sim0.8$ \citep{zahid.11.deep2}, $z\sim2.2$ \citep{steidel.14.mosfire, sanders.14.mzr,erb.06.mzr}, and $z\sim3.1$ \citep{mannucci.09.mzr}. We recall that these observed relations are originally obtained using different calibrations and the systematic uncertainty between different metallicity diagnostics could be up to 0.7 dex \citep{kewley.ellison.08}. To illustrate this point, we also convert all the observed relation to the N2 calibration from \citet[][PP04 hereafter]{pp04} unless their original data are already presented using this calibration. For \citet{tremonti.04.sdss} and \citet{zahid.11.deep2}, we do the conversion following the formula from \citet[][Table 3 therein]{kewley.ellison.08}. In either case, we present both their original relations and the converted relations using PP04 N2 calibration in Figure~\ref{fig:MZRGasObsComp}. At $z\sim2.2$, the observed relations are at already presented in PP04 N2 calibration \citep[e.g.][]{steidel.14.mosfire,sanders.14.mzr,erb.06.mzr}. \citet{mannucci.09.mzr} adopted a very different metallicity calibration, which is established using $z\gtrsim3$ galaxy samples only. Figure 5 in \citet{mannucci.09.mzr} suggests that the MZR evolves by $\sim0.4$ dex from $z\sim3.1$ to $z\sim2.2$. Motivated by their results, we also move the $z\sim2.2$ MZR from \citet{erb.06.mzr} downward by 0.4 dex for a comparison (lower right panel in Figure \ref{fig:MZRGasObsComp}).

In general, our simulations are in reasonable agreement with these observations in a broad range of stellar mass at $z=0$--3, especially when the observed relations are in their original forms. We emphasize that the empirical calibrations developed from the local universe are not necessarily valid for high-redshift galaxies \citep[e.g.][]{steidel.14.mosfire,kewley.15.z}. Given the large systematic uncertainties, we do not provide a detailed quantitative discussion of the discrepancies between our simulations and observations. Our results on the evolution of the MZR in Section \ref{sec:mzrevolve} are predictions that can be tested more accurately as our understanding of the observational systematic uncertainties improves.

In Figure~\ref{fig:MZRGasTheComp}, we also compare the MZR from our simulations with other cosmological simulations and semi-analytic models. We compare our results with two other simulations, \citet[][red lines]{torrey.14.illustris} and \citet[][green lines]{dave.11.mzr}, and three semi-analytic models from \citet[][the Lu model, magenta; the Somerville model, cyan; the Croton model, yellow]{lu.13.sam}. These models adopt ``sub-grid'' empirical models of galactic winds and stellar feedback, which couple some fraction of energy and/or momentum from SNe to the gas, and force certain amount of the gas to escape the galaxy. Note that the metal yields and solar abundance used in different works are not exactly the same, we renormalize all the $z=0$ MZR to ${\rm 12+\log(O/H)=8.9}$ at $\Ms=10^{10}~\Msun$ for comparison. At $z=0$, \citet{torrey.14.illustris} and the Lu model show steeper slopes at the low-mass end, due to the low metal retention efficiency in low-mass galaxies, a consequence of invoking strong outflows to suppress star formation in these galaxies\footnote{This can be simply illustrated using the ``leaky box'' model \citep[e.g.][]{schmidt.63.chem}. Assuming the outflow rate is proportional to the star formation rate ($\dot{M}_{\rm out}=\eta\cdot {\rm SFR}$, where $\eta$ is the mass loading factor), the metallicity is inversely proportional to $1+\eta$. Low-mass galaxies are very efficient in driving outflows and thus have high mass loading factors compared to massive galaxies. In SAMs and some simulations with ``sub-grid'' feedback models, it is often assumed that either the metals are well mixed in the system or that the outflowing gas has a metallicity comparable to the metallicity in the ISM. As a consequence, low-mass galaxies tend to lose a large fraction not only of their gas but also of their metals, and therefore to end up with very low metallicities.}. Some models predict higher metallicities at the most massive end. Furthermore, these models show significant discrepancies at $z\gtrsim2$. Our simulations predict much stronger evolution of MZR from $z=3$--0 than any other models. Particularly, the Somerville model and the Croton model predict inverse evolution trends -- the gas-phase metallicity decreases at lower redshifts at fixed stellar mass -- in contrast with observations and other models. We recall that although these models are tuned to match the observed stellar mass function at $z=0$, they tend to predict systematically higher stellar mass functions than the observed ones for $\Ms\lesssim10^{11}~\Msun$ at $z>0$ \citep{somerville.15.araa}, a consequence of the fact that galaxies in these models form {\it too} many stars at early time \citep[e.g.][]{dave.11.galaxy,sparre.14.illustris,fiacconi.15.argo}. In Section \ref{sec:sf}, we further explore how the different star formation histories between these models cause the discrepancies in the MZR at high redshifts.

\begin{figure}
\centering
\includegraphics[width=0.5\textwidth]{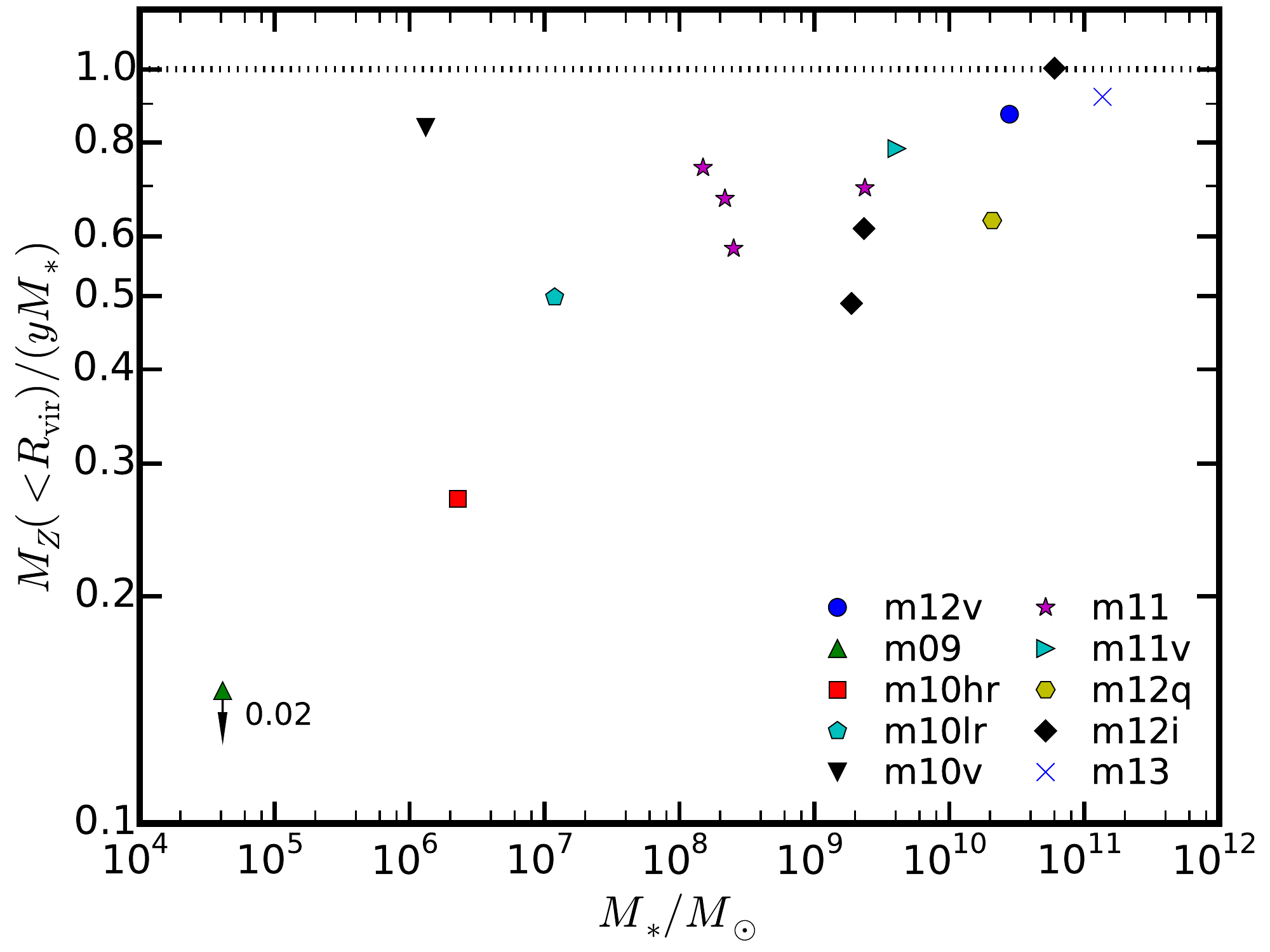}
\caption{Metal retention fraction for our simulated galaxies at $z=0$. $M_Z(<\Rvir)$ is the total amount of metals retained (in both gas and stars) within the virial radius. $y\Ms$ ($y$ is the mean effective yield) is the total metal mass produced by stars in the galaxies. The retained fraction of metal in the halo increases with stellar mass, from 30\% at $\Ms=10^6~\Msun$ to about unity at $\Ms>10^{10}~\Msun$. However, the ultra-faint dwarfs (e.g. {\bf m09}) are only able to retain 2\% of their metals in the halo.}
\label{fig:MetalLoss}
\end{figure}

\begin{figure*}
\centering
\includegraphics[width=1.0\textwidth]{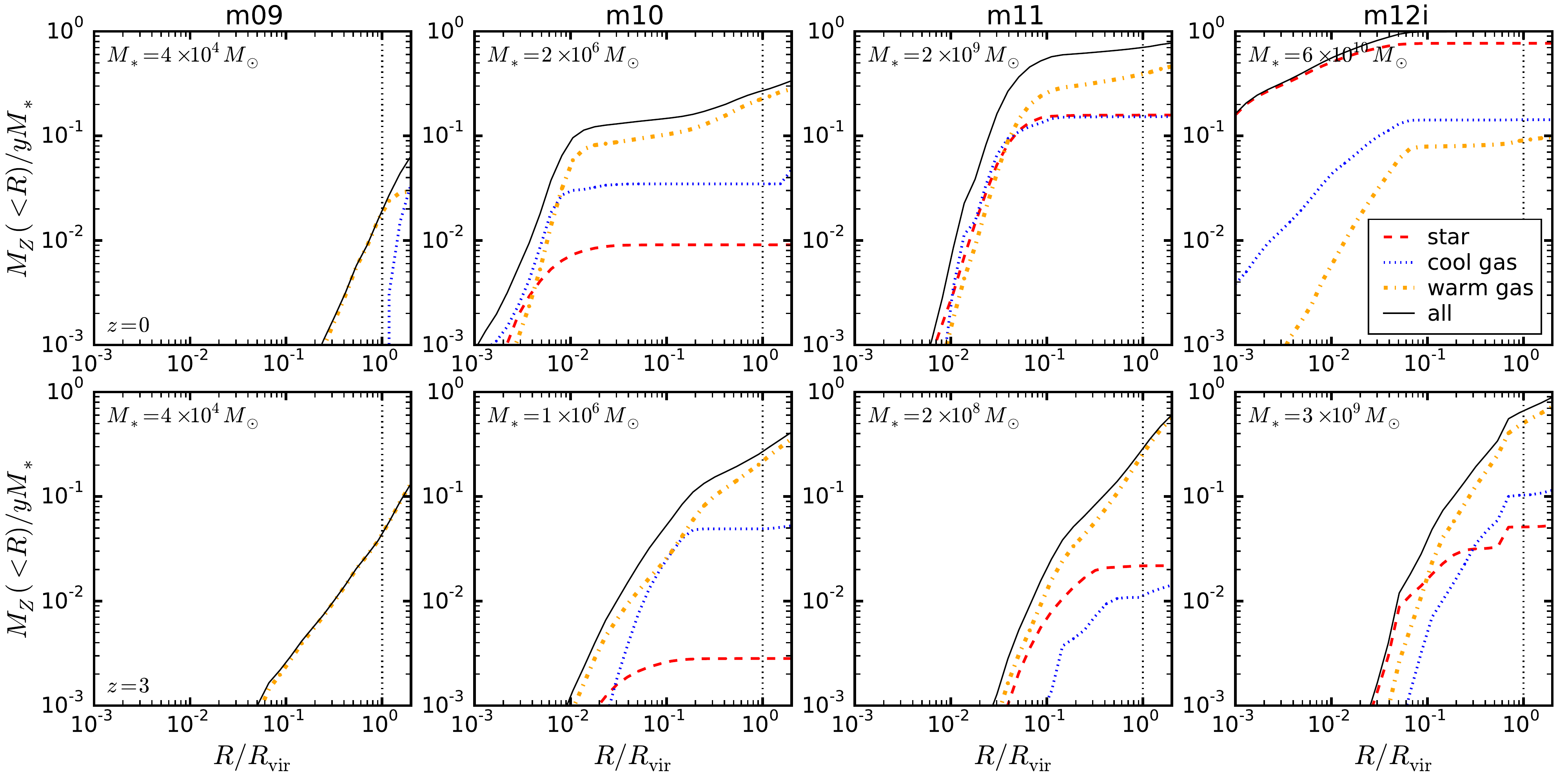}
\caption{Cumulative metal mass in selected simulated halos at $z=0$ (top) and $z=3$ (bottom), normalized by the total metal mass produced by stars ($y\Ms$). The red dashed, blue dotted, orange dash-dotted, and black solid lines show the metal mass in stars, cool gas ($T<10^4$ K), warm gas ($10^4~{\rm K}<T<4\times10^5$ K), and total, respectively. The stellar mass of each galaxy is indicated at the top left corner of each panel and the black dotted lines show the virial radius. At $z=0$, most of the metals in our more massive simulated galaxies such as {\bf m12i} are in stars and within 0.1 $\Rvir$ of the halo center, while in low-mass galaxies, the majority of metals are found in the warm CGM. In low-mass galaxies at $z=0$ and in high-redshift galaxies, a larger fraction of the metals are found at larger radii from the halo center, consistent with the fact that galactic outflows are more powerful in these systems.}
\label{fig:MetalPhase}
\end{figure*}

\begin{figure}
\centering
\includegraphics[width=0.5\textwidth]{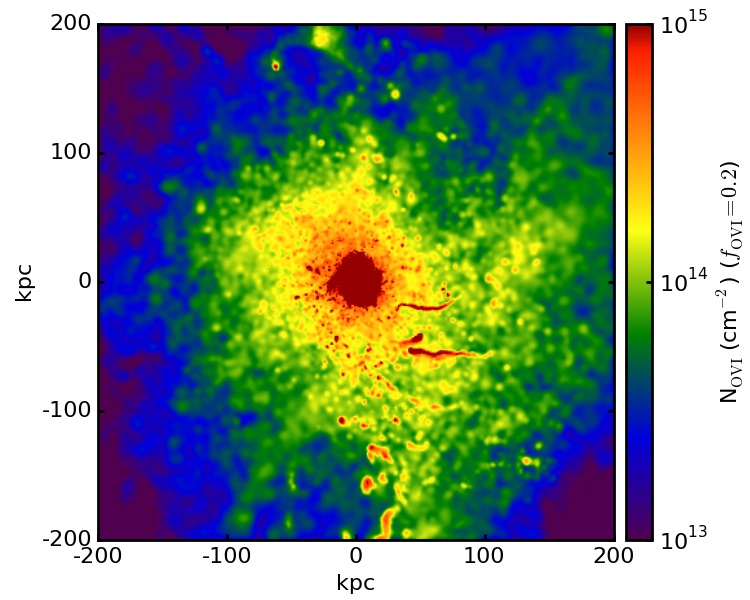}
\caption{O VI column density map for the m12i halo at $z=0$. We crudely assume that a fraction $f_{\rm OVI}=0.2$ of the oxygen is in O VI and only include warm and hot gas ($T>10^4$ K) in the halo. The characteristic $N_{\rm OVI}$ drops from $\sim10^{15}$ cm$^{-2}$ at impact parameter $b=20$ kpc from the central galaxy to $\sim10^{13.5}$ cm$^{-2}$ at $b=200$ kpc. The simulation agrees well with the O VI columns measured by COS-Halos \citet{tumlinson.11.cos} around low-redshift $\sim L^{*}$ star-forming galaxies at impact parameters $b<50$ kpc but appears to underestimate O VI columns by a factor of a few at larger impact parameters. Overall the agreement with observed O VI columns is reasonable given the uncertainties in ionization correction.}
\label{fig:CGMMetal}
\end{figure}

\begin{figure*}
\centering
\includegraphics[width=0.8\textwidth]{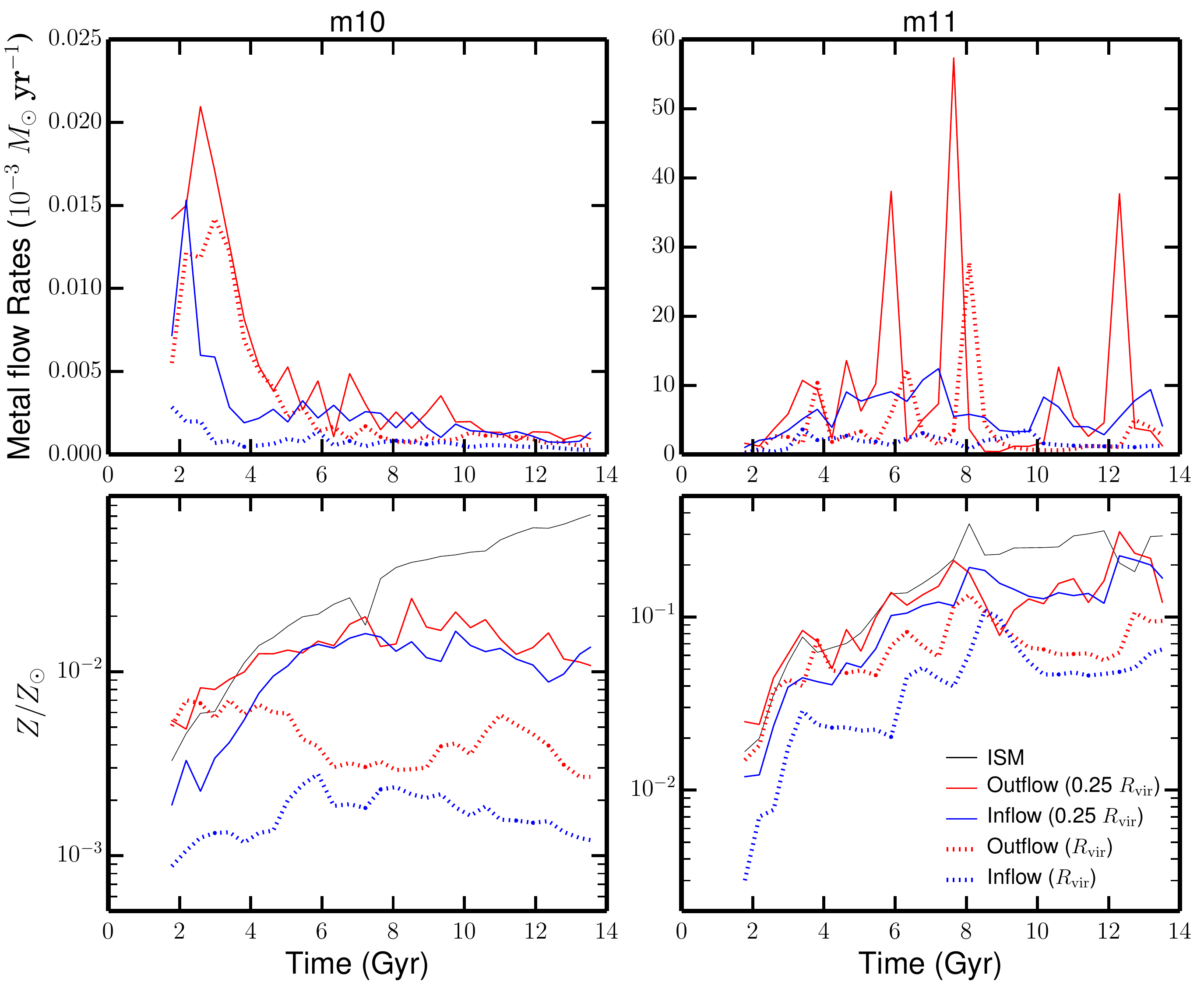}
\caption{{\it Upper}: Metal inflow (blue) and outflow rates (red) from $z=0$--4. Solid and dotted lines show the metal inflow/outflow rates measured at 0.25 $\Rvir$ and $\Rvir$, respectively. {\it Bottom}: Metallicities of inflowing/outflowing gas. The black line shows the metallicity of the ISM. All quantities are averaged over a time-scale of 400 Myr. Metals are efficiently ejected in fountains reaching 0.25 $\Rvir$, but they do not usually reach $\Rvir$ -- they are either deposited in the halo or recycled efficiently in galactic fountains. Outflowing gas that escapes from the halo at $\Rvir$ tends to be less enriched than the gas in the ISM.}
\label{fig:Outflow}
\end{figure*}

\begin{figure*}
\centering
\includegraphics[width=1.0\textwidth]{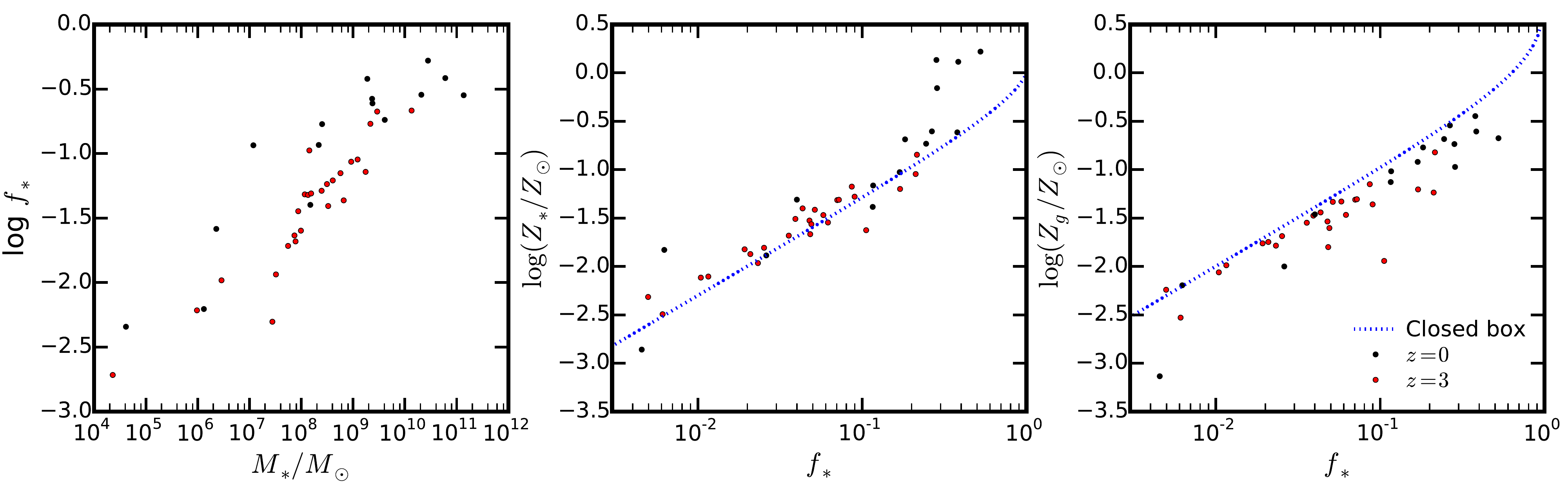}
\caption{{\it Left}: Stellar mass fraction $f_{\ast}=\Ms/(M_{\rm gas}+\Ms)$ as a function of stellar mass. $M_{\rm gas}$ here is the total gas mass in the {\it halo} (not only in the galaxy). {\it Middle}: Stellar metallicity $\Zs$ as a function of $f_{\ast}$. {\it Right}: Gas-phase metallicity $\Zg$ as a function of $f_{\ast}$. For consistency, $\Zg$ here is the average metallicity of {\it all} gas in the halo (including {\it both} the ISM {\it and} the halo gas). Black points and red points show the primary FIRE simulations at $z=0$ and $z=3$, respectively. Blue dotted lines show the simple ``closed box'' model predictions assuming an effective metal yield of $y=0.02$. The $z=0$ and $z=3$ galaxies share the same $\Zs$--$f_{\ast}$ and $\Zg$--$f_{\ast}$ relations, but the $f_{\ast}$--$\Ms$ relation evolves by $\sim0.5$ dex from $z=3$--0. This indicates that the evolution of the MZR is associated with the evolution of $f_{\ast}$ (at a fixed stellar mass) at different redshifts. The major offset between our simulations and the predictions of the ``closed box'' model is largely due to the fact that the metals are not perfectly mixed throughout the halo. Especially in massive galaxies, gas tends to be more metal-enriched in the central star-forming regions than in the outer halo, so stellar metallicities tend to be higher and gas-phase metallicities (including the halo gas) are lower than the predictions of the ``closed box'' model.}
\label{fig:CloseBox}
\end{figure*}

\section{Discussion}
\label{sec:discussion}
We showed above that the gas-phase and stellar MZR in our simulations agree broadly with available observations at different redshifts. We also found that our predictions diverge significantly from those of several large-volume cosmological hydrodynamical simulations and semi-analytic models. In this section, we explore the key factors that drive the shape and evolution of the MZR and discuss why our predictions differ from some other models.

\subsection{Where are the metals?}
Our simulations produce much higher metallicities for galaxies of stellar mass $\Ms<10^9~\Msun$ than \citet{torrey.14.illustris} and the Lu model in \citet{lu.13.sam}, indicating that our low-mass galaxies retain more metals compared to those models, despite the fact that these galaxies have high wind mass loading factors up to 100. To explicitly show this, we present in Figure~\ref{fig:MetalLoss} the metal mass fraction retained within $\Rvir$ as a function of stellar mass for the simulated sample at $z=0$. The numbers are obtained as follows. First, we estimate the effective yield $y$ for every simulation as the ratio between total metal mass (in both gas and stars) and the total stellar mass in the whole simulation volume. Then the metal retention fraction for a galaxy is simply the ratio between the total metal mass within the virial radius, $M_Z(<\Rvir)$, and $y\Ms$, where $\Ms$ is the galaxy stellar mass as defined in Section \ref{sec:halo}. Thus, $y\Ms$ represents the total amount of metal ever produced by the stars in the galaxy. As shown in Figure \ref{fig:MetalLoss}, the metal retention fraction generally increases with stellar mass. In our simulated sample, galaxies above $\Ms=10^{10.5}~\Msun$ are able to keep almost all metals they have produced. At much lower masses ($\Ms=10^6$--$10^7~\Msun$), they can still retain at least 30\% to a half of their metals within the halo. In contrast, the ultra-faint dwarf in our sample, {\bf m09} ($\Ms=4\times10^4~\Msun$), only retains 2\% of its metals within $\Rvir$ at $z=0$.

To quantify in more detail how metals are retained in galaxy halos, we show in Figure \ref{fig:MetalPhase} the cumulative metal retention fraction, as a function of radius, for different gas phases (cool gas with $T<10^4$ K and warm gas with $10^4~{\rm K}<T<4\times10^5$ K)\footnote{In our simulations, most of the diffuse ($n_{\rm H}<0.1$ cm$^{-3}$) gas has temperature $T>10^{4}$ K so a temperature cut at $T=10^{4}$ K also effectively separates ISM and CGM gas, justifying our approach of using gas with $T<10^{4}$ K to evaluate gas-phase ISM metallicities.}. At $z=0$ (top row), low-mass galaxies such as {\bf m10} ($\Ms=2\times10^6~\Msun$) have most of their metals in the warm CGM, while in massive galaxies like {\bf m12i} ($\Ms=6\times10^{10}~\Msun$), the majority of the metals are found in stars. 
This trend is qualitatively consistent with the empirical halo metal budget presented in \citet[][Fig. 6]{peeples.14.cos}. In most cases, we find that only a small fraction of the total metal mass is found in hotter ($T>4\times10^5$ K) gas. Our results are in contrast with the large-volume simulations of \citet{ford.15.cos} based on a parameterized galactic wind model, in which stars, ISM, and the cool CGM contain comparable metal masses for halos of mass similar to our m12i run.

Regarding the spatial distribution of metals, in {\bf m11} ($\Ms=2\times10^9~\Msun$), over 60\% of the metals are concentrated in the central 0.1 $\Rvir$ (mostly the galaxy) and only a small fraction ($\lesssim40\%$) of metals are in the circumgalactic medium (CGM) or lost into the IGM. In massive systems such as {\bf m12i} ($\Ms=6\times10^{10}~\Msun$), almost all the metals are in the central 0.1 $\Rvir$. In low-mass galaxies like {\bf m10} ($\Ms=2\times10^6~\Msun$), metals are more evenly distributed among the galaxy, the CGM, and the IGM. In ultra-faint dwarfs like {\bf m09} ($\Ms=4\times10^4~\Msun$), most of the metals it has ever produced are lost in the IGM by $z=0$. This is consistent with the fact that outflows in low-mass galaxies are more efficient (they have much higher mass loading factor) and can propagate more easily to large radii than in massive systems \citep{muratov.15.outflow}. It also shows that the metals are far from well mixed in the halo of more massive galaxies with stellar mass $\Ms>10^9~\Msun$ (or in terms of halo mass $\Mvir>10^{11}~\Msun$).

For a comparison, we also show the cumulative metal distribution for the progenitors of these galaxies at $z=3$ (the bottom panel in Figure \ref{fig:MetalPhase}). Similar to $z=0$, a significant fraction of metals are still retained in $\Rvir$ at $z=3$, although metals are more uniformly distributed from the centre to a few virial radii. These galaxies have much lower mass than their low-redshift decedents, and thus they are more efficient in driving gas outflows from star-forming regions throughout the halo.

\subsection{Circum-galactic O VI}
Although this paper is primarily focused on the metallicity of gas and stars inside galaxies, it is useful to check whether our simulations are consistent with observed CGM metal absorption. In addition to the overall metal budget discussed above, the COS-Halos program has provided useful measurements of O VI absorption around $\sim L^{*}$ galaxies at $z\approx0.1-0.4$ \citep[][]{tumlinson.11.cos}. Figure \ref{fig:CGMMetal} shows the O VI column density map around our m12i simulated halo at $z=0$. For this comparison, we assume that a fraction $f_{\rm OVI}=0.2$ of the oxygen is in O VI and only include warm and hot gas ($T>10^4$ K) in the halo. $f_{\rm OVI}=0.2$ is the maximum expected if the oxygen is in collisional ionization equilibrium, though it is possible that OVI is also photoionized and/or subject to non-equilibrium effects \citep[e.g.,][]{oppenheimer.09.ovi,oppenheimer.13.ovi} so that this ionization fraction is not a strict upper limit. The figure shows that for m12i the characteristic $N_{\rm OVI}$ drops from $\sim10^{15}$ cm$^{-2}$ at impact parameter $b=20$ kpc from the central galaxy to $\sim10^{13.5}$ cm$^{-2}$ at $b=200$ kpc. The simulation agrees well with the O VI columns measured by \citet{tumlinson.11.cos} around low-redshift $\sim L^{*}$ star-forming galaxies at impact parameters $b<50$ kpc but appears to underestimate O VI columns by a factor of a few at larger impact parameters. Overall the agreement with observed O VI columns is reasonable given the uncertainties in ionization correction. More systematic and detailed comparisons of CGM metal statistics from the FIRE simulations with observations will be reported in future papers (Hafen et al., in preparation).

\subsection{Metal outflows, inflows, and recycling}
SAMs and large-volume cosmological simulations require ``sub-grid'' models of galactic winds, which often incorporate fairly crude approximations. In this subsection, we further examine the metal inflow and outflow rates and the metallicities of gas inflows and outflows in our simulations and compare with the assumptions of common ``sub-grid'' models.

We follow \citet{cafg.11.flow} and \citet{muratov.15.outflow} and define the gas outflow rates, metal outflow rates, and metallicities of outflow gas as 
\be
  \frac{\partial M}{\partial t} = \sum_i \vec{v} \cdot \frac{\vec{r}}{|r|} M_i / \mathrm{d}L, 
\ee
\be
  \frac{\partial M_{\rm metal}}{\partial t} = \sum_i \vec{v} \cdot \frac{\vec{r}}{|r|} Z_i~M_i / \mathrm{d}L,
\ee
\be
  Z_{\rm outflow} = \frac{\partial M_{\rm metal}}{\partial t} / \frac{\partial M}{\partial t},
\ee
where $M_i$ and $Z_i$ are the mass and metallicity of the $i^{\rm th}$ gas particle within the shell of thickness d$L=0.1~\Rvir$ with radial velocity outwards $\vec{v} \cdot \frac{\vec{r}}{|r|} > 0$. The inflow rates and inflow metallicities are defined in the same way but for gas particles with inward radial velocity $\vec{v} \cdot \frac{\vec{r}}{|r|} < 0$. The upper panels in Figure \ref{fig:Outflow} show the metal inflow/outflow rates at $0.25~\Rvir$ (blue/red solid lines) and at $\Rvir$ (blue/red dotted lines) for our {\bf m10} (left) and {\bf m11} (right) simulations. We average the inflow/outflow rates on a time-scale of 400 Myr. In either case, the net metal outflow rates are considerably lower at $\Rvir$ than at $0.25~\Rvir$, indicating that the metals are either deposited in the halo or returned back to the ISM. At high redshifts, metals ejected in outflows can be more easily driven to $\Rvir$ than at low redshifts. At 0.25 $\Rvir$, metal inflow rates are comparable to metal outflow rates, suggesting a high efficiency of metal recycling. The lower panels in Figure~\ref{fig:Outflow} show the average metallicities of inflows and outflows at both $0.25~\Rvir$ and at $\Rvir$, as compared to the metallicity of the ISM (black solid lines). The outflow metallicities are much lower at $\Rvir$ than at $0.25~\Rvir$ (and even more so than in the ISM), because outflowing gas sweeps up and mixes with more metal-poor gas in the halo as it propagates outwards. This is particularly important for low-mass galaxies, such as {\bf m10} ($\Ms=2\times10^6~\Msun$), which can have wind mass loading factors up to $\sim100$, yet retain a large fraction of the metals they produced in their halos.

Our analysis calls into question a number of assumptions and approximations often adopted in analytic, semi-analytic, and large-volume cosmological hydrodynamic models of galaxy formation. First of all, unlike often assumed in analytic and semi-analytic models, metals are generally not well-mixed in galaxy halos (e.g. Figure \ref{fig:MetalPhase}). In particular, in many ``sub-grid'' galactic wind models, wind gas is assumed to have a metallicity directly related to the ISM metallicity \citep[e.g.][]{dave.11.galaxy,torrey.14.illustris}, an assumption that oversimplifies the complex mass and metal loading that takes places in our more explicit simulations. Our simulations also indicate that metal re-accretion onto galaxies (recycling) is important on small scales, an effect which is not well captured in semi-analytic models and in ``sub-grid'' models that either assume that the ejected gas never returns to the galaxy, or which ignore hydrodynamical interactions between the wind and the gas close to the galaxy.

Recently, \citet{lu.15.3model} compared three different SAM feedback models --- one including only gas ejection, one including both gas ejection and recycling, and the other including a model of ``preventive'' feedback. \citet{lu.15.3model} found that none of these models could {\it simultaneously} reproduce the MZR, the distribution of metals in different phases, and the SFR observed at $z=0$--3. This finding is consistent with the picture suggested by our high-resolution simulations that the chemical evolution of galaxies is a complex process and that it is necessary to self-consistently model galaxy-halo interactions in order to capture it faithfully. It is encouraging that our cosmological simulations with explicit stellar feedback and hydrodynamical interactions tracked at all times appear to produce a low-mass-end slope of the MZR that is closer to observations than most previous models, without the need for parameter tuning. Our results are broadly consistent with those of \citet{brook.14.sim}, who also highlighted the importance of metal mixing with the CGM and recycling for explaining the MZR. The simulations of \citet{brook.14.sim} also provide a fair match to the observed MZR at $z=0$--3 \citep{obreja.14.magicc}).

\subsection{Why the MZR evolves with redshift?}
\label{sec:sf}
Another major difference between our simulations and other theoretical work is we predict much stronger evolution of the MZR from $z=3$--0 (e.g. the stellar metallicity increases by 0.5 dex at fixed stellar mass, see Figure \ref{fig:ZeroPoint}). Observations and some theoretical models suggest a fundamental metallicity relation (FMR) between stellar mass, star formation rate, and metallicity that holds for star-forming galaxies both in the local universe and at high redshifts \citep[e.g.][]{mannucci.10.fmr,mannucci.11.fmr,lilly.13.model,obreja.14.magicc,cullen.14.mzr,zahid.14.fmr}. Motivated by these results, we attempt to qualitatively illustrate what might be the primary factor that drives the evolution of MZR in this section. We start by reviewing the simplest chemical evolution model, i.e., the ``closed box'' model, which predicts the stellar and gas-phase metallicities as a function of stellar mass fraction, $f_{\ast} = \Ms/(M_{\rm gas}+\Ms)$ as the following
\ba
  Z_{\ast} & = & y ~ \left[ \frac{1-f_{\ast}}{f_{\ast}} \ln(1 - f_{\ast}) + 1 \right], \\
  Z_g & = & - y ~ \ln (1 - f_{\ast}),
\ea
where $y$ is the effective metal yield \citep[e.g.][]{schmidt.63.chem,talbot.71.chem,searle.72.chem}. The parameter $f_{\ast}$ describes the fraction of baryons that have been turned into stars, and $1-f_{\ast}$ is the ``gas fraction''. In Figure~\ref{fig:CloseBox}, we show the relation between stellar and gas-phase metallicities and $f_{\ast}$, respectively (the middle and right panels), for our {\bf mxx} series simulations at $z=0$ and $z=3$ (black and red points). We emphasize that we account for {\it both} the halo gas {\it and} the ISM in the total gas mass when calculating $f_{\ast}$, since halo gas is actively involved in supplying star formation and metal exchange in most cases. For consistency, the gas-phase metallicities shown in Figure \ref{fig:CloseBox} is the average metallicity of {\it all} gas in the halo. For illustrative purpose, we also show the simple predictions from the ``closed box'' model, assuming an effective metal yield of $y=0.02$ (blue dotted lines in Figure \ref{fig:CloseBox}). 

The simulated data at $z=0$ and $z=3$ overlap with each other in the $\Zs$--$f_{\ast}$ and $\Zg$--$f_{\ast}$ diagrams. In the left panel of Figure \ref{fig:CloseBox}, we also show the relation between $f_{\ast}$ and $\Ms$ for these galaxies at both redshifts. There is a systematic offset ($\sim$0.5 dex) in the $f_{\ast}$--$\Ms$ relation between galaxies at $z=0$ and 3. Note that in the limit of $f_{\ast}\ll1$, one has $\Zs,\Zg \propto f_{\ast}$. Therefore, the 0.5 dex offset in $f_{\ast}$--$\Ms$ relation propagates to the 0.5 dex evolution of the MZR from $z=3$ to 0. This suggests that the evolution of the MZR is associated with the evolution of $f_{\ast}$ (at a fixed stellar mass) within the {\it halo} at different redshifts, providing a first hint of a universal metallicity relation between stellar mass, gas mass, and metallicities \citep[cf.][for observational evidences]{bothwell.13.fmr,zahid.14.fmr}. In simulations with ``sub-grid'' feedback models and semi-analytic models, where the $z=0$ stellar mass functions are tuned to match observations, galaxies tend to form a large fraction of their stars at high redshift and therefore their evolution is weaker at lower redshift \citep[e.g.][]{somerville.15.araa}, as opposed to observations and our simulations. In other words, these models produce higher $f_{\ast}$ than our simulations at fixed stellar mass at $z>0$ and an $f_{\ast}$--$\Ms$ relation barely evolving from $z=3$--0. Therefore, galaxies in those models are more metal-enriched at high redshifts and the evolution of the MZR is weaker than our simulations.

Our simulations are qualitatively consistent with the simple ``closed box'' predictions applied to {\it halo} quantities\footnote{We emphasize that in the analog of Figure \ref{fig:CloseBox} where we measure $f_{\ast}$ using {\it only} the gas in the galaxy (i.e., excluding the halo gas), all the galaxies are well below the predictions of the closed box model and there is no well-defined relation, indicating that galaxies themselves are far from closed boxes. This suggests the necessity of accounting for halo gas as reservoirs in galaxy evolution.}. This is not unreasonable because a large fraction (order unity) of metals are retained within the virial radius at both redshifts (see e.g., Figure \ref{fig:MetalPhase}). However, we emphasize that one should not think our simulated galaxies are closed boxes, because the metals are not perfectly well-mixed in the galactic halo. This explains the major offset between the ``closed box'' model and our simulations (middle and right panels in Figure \ref{fig:CloseBox}), especially in the most massive systems where this effect is stronger. Since gas in the centre of the galaxy tends to be more metal-enriched than gas in the outer halo and stars preferentially form in the central region, stellar metallicities tend to be higher and the gas-phase metallicities (including the halo gas) are lower than the predictions of the closed box model (applied to halo quantities). The mixing of metals is very complex and associated with galactic fountains on different scales. Although the ``closed box'' model gives a natural relation between stellar mass, gas mass, and the metallicities, the parameterization of a universal metallicity relation for galactic quantities (i.e., excluding the halo) is more complicated than the simple model. This is worth further investigation in more detail in future work.

\section{Conclusion}
\label{sec:conclusion}
We use a series of high-resolution cosmological zoom-in simulations spanning halo masses $10^9$--$10^{13}~\Msun$ and stellar masses $10^4$--$10^{11}~\Msun$ at $z=0$ from the FIRE project to study the galaxy mass--metallicity relations at $z=0$--6. These simulations include explicit models of multi-phase interstellar medium, star formation, and stellar feedback. As has been shown in previous papers, these simulations successfully reproduce many observed galaxy properties, including the stellar mass--halo mass relation, star-forming main sequence, the Kennicutt-Schmidt law, star formation histories, etc., for a wide range of galaxies at many redshifts \citep{hopkins.14.fire}. These simulations also predict reasonable covering fractions of neutral hydrogen in the halos of $z=2$--3 LBGs \citep{cafg.14.fire} and self-consistently generate galactic winds with velocities and mass loading factors broadly consistent with observational requirements \citep{muratov.15.outflow}. These simulations adopt ``standard'' stellar population models and metal yield tables from Type-I and Type-II supernovae and stellar winds, following species-by-species for 11 separately tracked elements. Our key conclusions include the following.

(i) The simulations predict galaxy mass--metallicity relations that agree reasonably well with a number of observations from $z=0$--3 for a broad range of stellar masses. Both gas-phase and stellar metallicities evolve monotonically from $z=0$--6, with higher metal abundance at low redshifts at fixed stellar mass. The best linear fits of the MZR for our simulated galaxies as a function of redshift are $\log(\Zg/\Zsun) = {\rm 12 + \log(O/H) - 9.0} = 0.35~[\log(\Ms/\Msun)-10] + 0.93 \exp(-0.43z) - 1.05$ and $\log(\Zs/\Zsun) = {\rm [Fe/H]} + 0.2 = 0.40~[\log(\Ms/\Msun)-10] + 0.67 \exp(-0.50z) - 1.04$, for gas-phase metallicity and stellar metallicity, respectively. We emphasize that the normalizations may have systematic uncertainties that originate from the SNe rates, yield tables, and solar abundance we adopt, but the evolution of the MZR is robust to these uncertainties.

(ii) The stellar MZR becomes flat around $\Ms\sim10^{11}~\Msun$ since $z=0$, because the most massive galaxies in our simulations evolve via mergers and accretion of satellites rather than {\it in situ} star formation at low redshifts. Therefore, the stellar metallicity does not increase despite the fact that the stellar mass grows considerably. We do not see the flatness in the gas-phase MZR at the high-mass end seen in observations, because gas continues to be enriched by non-negligible star formation. This apparent discrepancy may be due to the more limited resolution in our m13 run or to the lack of AGN feedback in our simulations. AGN might be required to quench star formation below $z\sim1$ in such massive galaxies.

(iii) The evolution of MZR is associated with the evolution of the gas/stellar mass fraction within the inner halo (not just inside the galaxy effective radius) at different redshifts. This provides a first hint of a universal metallicity relation between stellar mass, gas mass, and metallicities, but its parameterization for galactic quantities (as opposed to for halo quantities, which behave more like a closed box) is much more complicated than simple analytic models. We will investigate this in more detail in future work.

(iv) Galaxies above $10^6~\Msun$ can retain a large fraction of their metals in the halo even up to $z=3$. The net metal outflow rates near the virial radius are always lower than those near the galaxy, indicating that the metals either get deposited in the halo or return back to the ISM. The high metal inflow rates and the high metallicity of inflowing gas at 0.25 $\Rvir$ suggest a high efficiency of metal recycling (a finding that we have confirmed using particle tracking; Angl\'es-Alc\'azar et al., in prep.). On average, the outflows at outer radii are much less metal-enriched than those at the inner radius. This effect helps resolve the tension between the need for strong gas outflows and high metal retention fractions in low-mass galaxies.

(v) These differential recycling and metal retention effects are not properly accounted for in most semi-analytic and early generation of ``sub-grid'' feedback models that are popular in cosmological simulations. As a result, these simplified models cannot simultaneously reproduce the galaxy mass function and the slope and redshift evolution of the MZR. By explicitly resolving the ``missing physics'' in these models, we reconcile the long-standing discrepancy, and provide a clear way forward to improve the sub-grid and semi-analytic models.

Nevertheless, our simulations are still limited in sample size. In the near future, we will expand our simulations to include more dwarf galaxies covering halo mass from $M_{\rm halo}=10^8$--$10^{11}~\Msun$ and to enlarge our sample at the most massive end to better understand whether the flattening of the MZR is real and what drives the flatness. This may depend critically on AGN feedback. We will provide quantitative analysis on metal outflow rates, outflow metallicities, metal recycling, and their relation with galaxy properties in future work (Muratov et al., in preparation; Angl{\'e}s-Alc{\'a}zar et al, in preparation).

\section*{Acknowledgments}
We thank Daniel Angl{\'e}s-Alc{\'a}zar, Yu Lu, Evan Kirby, Paul Torrey, Andrew Wetzel, and many friends for helpful discussion and useful comments on this paper. We also thank Jabran Zahid, Robert Yates, Chris Brook, and many others for their discussion after the first draft of this paper was submitted to arXiv.
The simulations used in this paper were run on XSEDE computational resources (allocations TG-AST120025, TG-AST130039, and TG-AST140023).
Support for PFH was provided by the Gordon and Betty Moore Foundation through Grant 776 to the Caltech Moore Center for Theoretical Cosmology and Physics, by the Alfred P. Sloan Foundation through Sloan Research Fellowship BR2014-022, and by NSF through grant AST-1411920. 
CAFG was supported by NSF through grant AST-1412836, by NASA through grant NNX15AB22G, and by Northwestern University funds.
DK was supported by NSF grant AST-1412153 and UC San Diego funds. 
EQ was supported by NASA ATP grant 12-APT12-0183, a Simons Investigator award from the Simons Foundation, the David and Lucile Packard Foundation, and the Thomas Alison Schneider Chair in Physics at UC Berkeley.

\bibliography{}

\appendix

\begin{figure*}
\centering
\begin{tabular}{ll}
\includegraphics[width=0.35\textwidth]{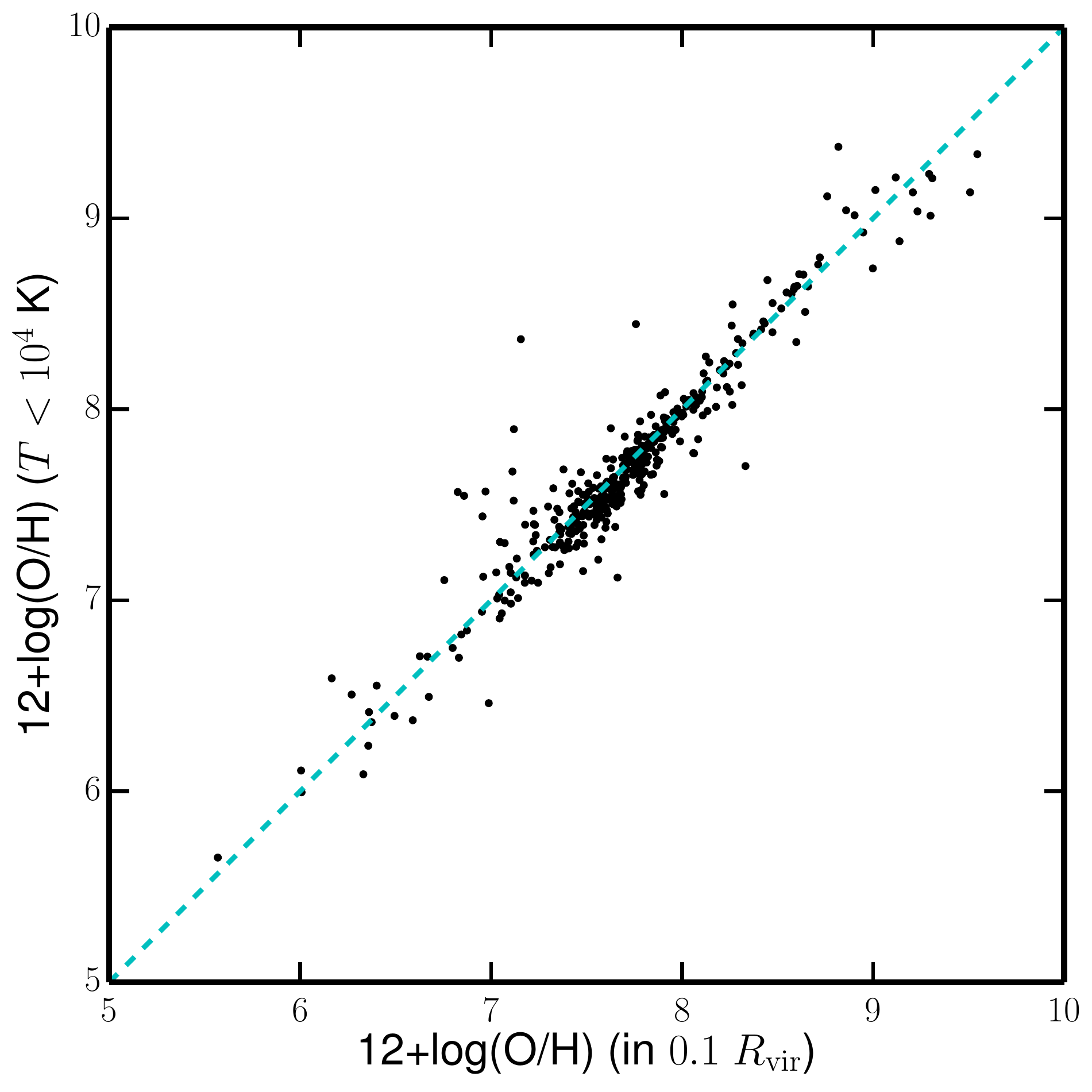} &
\includegraphics[width=0.35\textwidth]{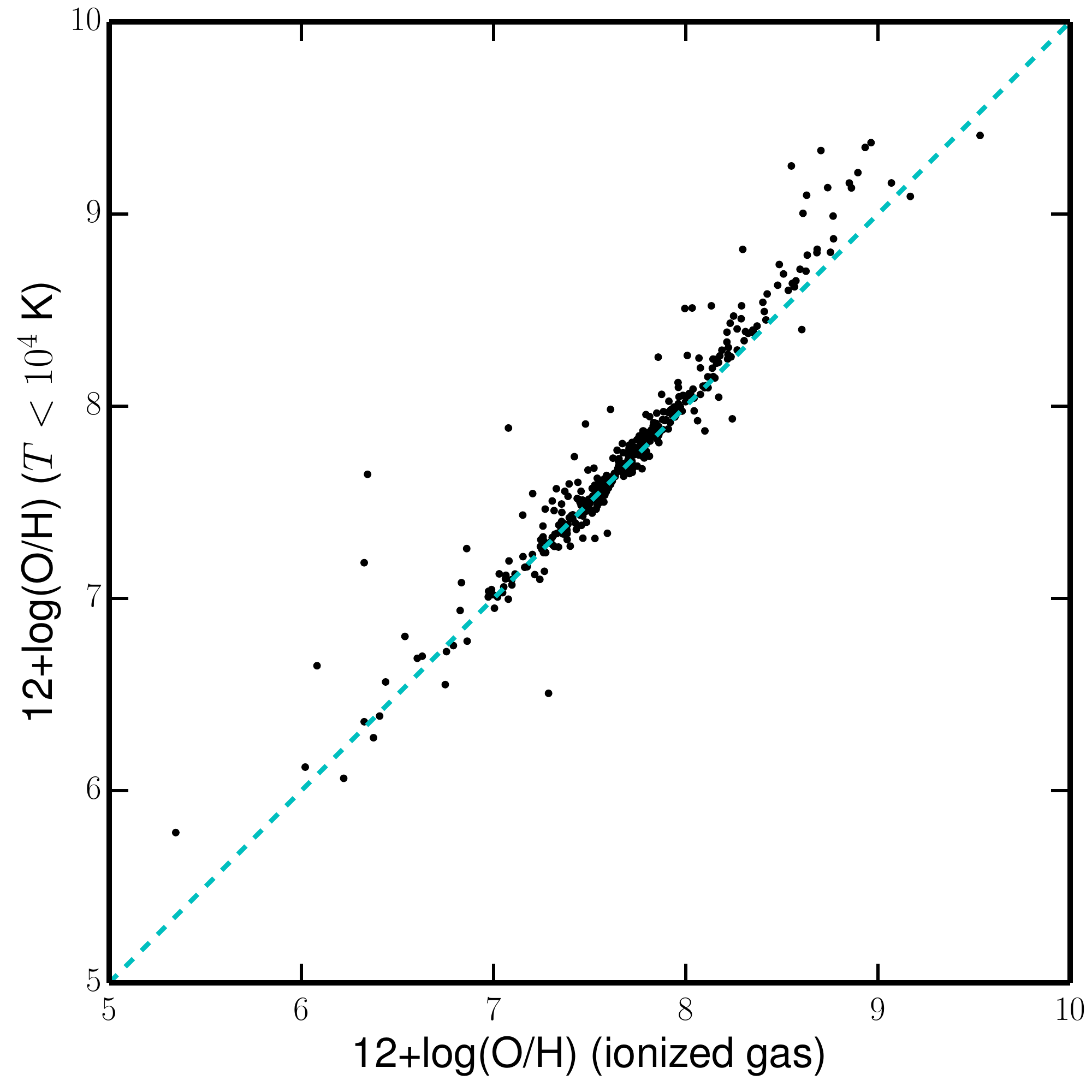}
\end{tabular}
\caption{Gas-phase oxygen abundances in different definitions. {\it Left:} The relation of gas oxygen abundances between definition (1) the average metallicity of all gas particles below $10^4$ K and (2) the average metallicity of all gas particles within 0.1 $\Rvir$. {\it Right}: The relation of gas oxygen abundances between definition (1) and (3) the average metallicity of all gas particles with temperature between 7,000--15,000 K and density above $0.5~\cm^{-3}$. The cyan dashed lines show the $y=x$ relation. The black points show all the data presented in Figure \ref{fig:MZRGasAllz}. Different definitions agree well, and have {\it no} qualitative effect on any of our conclusions. Most of the ``outliers'' are caused by transient, stochastic time variability.}
\label{fig:ZgasCalib}
\end{figure*}

\begin{figure*}
\centering
\includegraphics[width=0.75\textwidth]{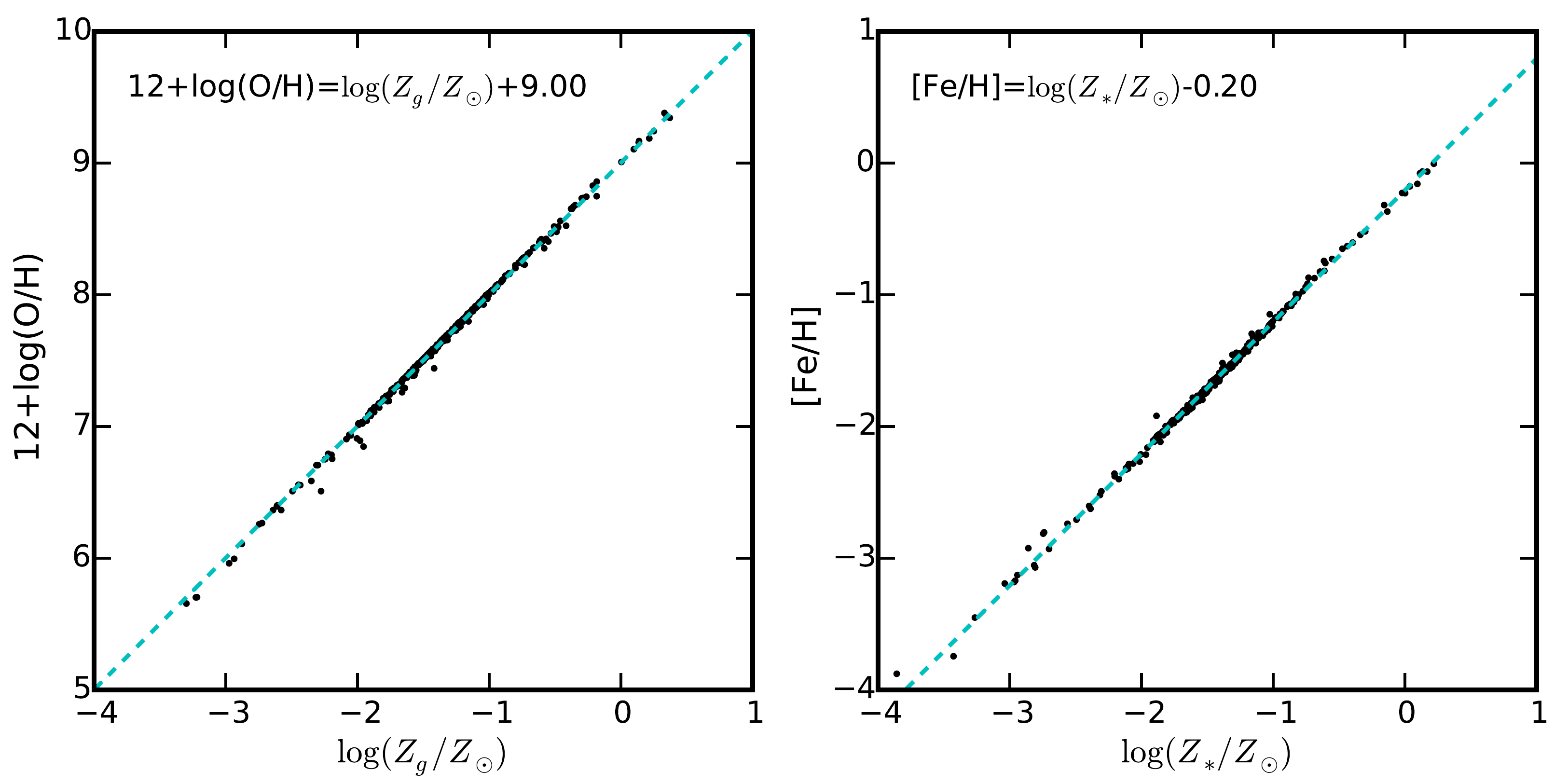}
\caption{Relations between different forms of metallicities. {\it Left}: Gas-phase oxygen abundance $\rm 12+\log(O/H)$ vs. gas-phase metallicity (mass fraction of all metals) $\Zg$. {\it Right}: Stellar iron abundance [Fe/H] vs. stellar metallicity $\Zs$. Black dots collect all the data points presented in this work. The cyan lines represent the best fits of these relations with slope unity. These definitions give essentially identical results, and are equivalent, for all of our results in this paper.}
\label{fig:ZCalib}
\end{figure*}

\section{Different Definitions of Gas-phase Metallicity}
\label{sec:zdefine}
In this work, the gas-phase metallicity is defined as the mass-weighted average metallicity of all gas particles below $10^4$ K, which we refer as the ISM gas. In principle, there are many alternative approaches to define gas-phase metallicities. In this section, we discuss three definitions and compare them with each other: (1) the average metallicity of all gas particles below $10^4$ K in the galaxy (our default definition), (2) the average metallicity of all gas particles within 0.1 $\Rvir$, and (3) the average metallicity of all gas particles with temperature between 7,000--15,000 K and density above $0.5~\cm^{-3}$. In Figure \ref{fig:ZgasCalib}, we compare definition (1) and (2) in the left panel and (1) and (3) in the right panel for all galaxies presented in Figure \ref{fig:MZRGasAllz}.

Definition (1) is designed to automatically select all the warm ionized gas and cold neutral gas (the ISM), definition (2) aims to pick the gas in the star-forming regions, and definition (3) is observationally motivated to select the nebular gas which produce the strong nebular emission lines in star-forming galaxies. In general, these definitions are consistent with each other. Most of the galaxies lie very close to the $y=x$ relation in each panel of Figure \ref{fig:ZgasCalib}. However, there are a few outliers in these diagrams. Definition (2) can be problematic in merging systems, where the halo centre may deviate far from the stellar bulk and thus 0.1 $\Rvir$ does not necessarily probe the star-forming region. Definition (3) is largely affected by abundance variance between gas particles, since there are usually not many gas particles at any single instant that meet the temperature and density criteria. However, a time-averaged version of definition (3) removes most of the outliers. Therefore, we argue that our default definition is more adaptive and flexible than other definitions.

\section{Metallicities in Different Forms}
\label{sec:zcalib}
In this work, we primarily use $\rm 12+\log(O/H)$ and $\Zs$ to present gas-phase metallicity and stellar metallicity, respectively. In the literature, gas-phase metallicity and stellar metallicity are sometimes presented in terms of $\Zg$ and [Fe/H]. Therefore, we also provide the conversion between these different forms of metallicities for comparison. We emphasize these conversions are obtained from our simulations only and there are systematic uncertainties originating from the uncertain relative metal yields between species and solar abundances we adopt.

In Figure \ref{fig:ZCalib}, we show the relations between $\rm 12+\log(O/H)$ and $\log(\Zg/\Zsun)$ (left panel) and the relation between [Fe/H] and $\log(\Zs/\Zsun)$ (right panel), where we adopt a solar metallicity $\Zsun=0.02$ and a solar iron abundance of 0.00173, both in mass fraction. In both panels, we collect data of all the simulated galaxies at all epochs we present earlier in this paper. Both relations are extremely tight and have slope unity, which ensures the validity, at least to the first order, to use either quantity to represent metallicities interchangeably. The best fits for our simulations are $\rm 12+\log(O/H)=log(\Zg/\Zsun)+9.0$ and ${\rm [Fe/H]}=\log(\Zs/\Zsun)-0.20$. We emphasize that there relations may suffer from systematic uncertainties that originate from: {\bf (1)} Type-II and Type-I SNe rates, {\bf (2)} metal yields of tracked species from different channels, and {\bf (3)} the solar abundances we adopt in our simulations.

\label{lastpage}

\end{document}